\documentclass[12pt]{article}
\usepackage{amsmath}
\usepackage{graphicx,psfrag,epsf}
\usepackage{enumerate}
\usepackage[longnamesfirst]{natbib}

\usepackage{amsfonts,amsmath,amssymb,amsthm,array,color,dsfont}
\usepackage{mathrsfs,mathtools,MnSymbol,multirow,rotating,setspace,soul,tabu,upgreek,ushort}	

\usepackage[english]{babel}
\usepackage [colorlinks=true,citecolor=black,linkcolor=black,urlcolor=blue]{hyperref}

\usepackage{algorithm,algpseudocode}

\renewcommand{\algorithmicrequire}{\textbf{Input:}\ }
\renewcommand{\algorithmicensure}{\textbf{Output:}\ }

\usepackage[inline]{enumitem}

\newcommand{\blind}{0}

\usepackage[affil-it]{authblk}
\newtheorem{assump}{Assumption}
\newtheorem{corol}{Corollary}

\newtheorem{lemma}{Lemma}
\newtheorem{alemma}{Lemma}[section]

\newtheorem{theorem}{Theorem}

\newtheorem{example}{Splitting Scheme}

\newtheorem*{excont}{Splitting Scheme \continuation}
\newcommand{\continuation}{??}
\newenvironment{continueexample}[1]
 {\renewcommand{\continuation}{\ref{#1}}\excont[continued]}
 {\endexcont}

\newcommand{\lowmathcal}[1]{{\operatorname{\text{\usefont{U}{BOONDOX-cal}{m}{n}#1}}}}

\newcommand{\Prob}{\mathbb{P}}

\newcommand{\E}{\mathbb{E}}
\newcommand{\V}{\mathbb{V}}
\newcommand{\convd}{\overset{d}{\rightarrow}}
\newcommand{\convp}{\overset{p}{\rightarrow}}

\newtheorem{innercustomgeneric}{\customgenericname}
\providecommand{\customgenericname}{}
\newcommand{\newcustomtheorem}[2]{%
  \newenvironment{#1}[1]
  {%
   \renewcommand\customgenericname{#2}%
   \renewcommand\theinnercustomgeneric{##1}%
   \innercustomgeneric
  }
  {\endinnercustomgeneric}
}

\newcustomtheorem{customassump}{Assumption}

\makeatletter
\newcommand*\bigcdot{{\mathpalette\bigcdot@{.75}}}
\newcommand*\bigcdot@[2]{\mathbin{\vcenter{\hbox{\scalebox{#2}{$\m@th#1\bullet$}}}}}
\makeatother

\usepackage[section]{placeins}
\usepackage{array,booktabs,multirow,placeins,rotating,tabu}
\makeatletter
\newcommand{\thickhline}{%
	\noalign {\ifnum 0=`}\fi \hrule height 1.5pt
	\futurelet \reserved@a \@xhline
}
\newcolumntype{"}{@{\hskip\tabcolsep\vrule width 1pt\hskip\tabcolsep}}
\makeatother

\usepackage{tabularx}

\newcommand\clearrow{\global\let\rowmac\relax}
\clearrow

\usepackage{accents}

\usepackage{xr}

\begin{document}

\def\spacingset#1{\renewcommand{\baselinestretch}%
{#1}\small\normalsize} \spacingset{1}


\if0\blind
{
  \title{\bf Nonparametric Estimation of Conditional Densities by Generalized Random Forests}
  \author{Federico Zincenko\thanks{
    I thank Gaurab Aryal, Marco Duarte, Seolah Kim, Tung Liu, and Ximing Wu for their helpful comments and discussions, as well as the participants at the conferences where this paper was presented: Midwest Econometrics Group Meeting (2022, MSU), SEA 92th Annual Meeting, UdeSA 12th Alumni Conference, WEIA 98th Annual Conference. I also thank Nianqing Liu and Yao Luo for kindly sharing their dataset. This work was completed utilizing the Holland Computing Center of the University of Nebraska, which receives support from the Nebraska Research Initiative.
    
    I did not receive any specific grant from funding agencies in the public, commercial, or not-for-profit sectors. I report that there are no competing interests to declare. \hspace{.2cm}}\\
    Deparment of Economics,\\
     University of Nebraska--Lincoln}
  \maketitle
} \fi

\if1\blind
{
  \bigskip
  \bigskip
  \bigskip
  \begin{center}
    {\LARGE\bf Nonparametric Estimation of Conditional Densities by Generalized Random Forests}
\end{center}
  \medskip
} \fi

\bigskip
\begin{abstract}
Considering a continuous random variable $Y$ together with a continuous random vector $X$, I propose a nonparametric estimator $\hat{f}(\cdot|x)$ for the conditional density of $Y$ given $X=x$. This estimator takes the form of an exponential series whose coefficients $\hat{\boldsymbol{\theta}}_{x}=(\hat{\theta}_{x,1}, \dots,\hat{\theta}_{x,J})$ are the solution of a system of nonlinear equations that depends on an estimator of the conditional expectation $\E[\boldsymbol{\phi}(Y)|X=x]$, where $\boldsymbol{\phi}$ is a $J$-dimensional vector of basis functions. The distinguishing feature of the proposed estimator is that $\E[\boldsymbol{\phi}(Y)|X=x]$ is estimated by generalized random forest (Athey, Tibshirani, and Wager, \emph{Annals of Statistics}, 2019), targeting the heterogeneity of $\hat{\boldsymbol{\theta}}_{x}$ across $x$. I show that $\hat{f}(\cdot|x)$ is uniformly consistent and asymptotically normal, allowing $J \rightarrow \infty$. I also provide a standard error formula to construct asymptotically valid confidence intervals. Results from Monte Carlo experiments are provided.
\end{abstract}

\noindent%
\vfill

\newpage
\spacingset{1.8} 

\section{Introduction}

The conditional density plays a key role in econometrics, as it provides a comprehensive description of how a random variable behaves when a specific value of another variable is given. Counterfactual exercises and policy recommendations are usually based on this description. To have a convincing assessment in this regard, it is crucial to rely on a conditional density estimator with good finite-sample performance and theoretical guarantees that rely on realistic assumptions.

Consider a continuous random vector $(Y , X)$ taking values on $[0,1] \times \mathscr{X}$, where $ \mathscr{X} \subset \mathbb{R}^{d}$ is a compact subset with nonempty interior and $d   \in \mathbb{N}$, and let $f(\cdot | x)$ denote the conditional density of $Y$ given $X=x$. In this setting, I propose a nonparametric estimator of $f(\cdot | x)$. Such an estimator is built by combining \cite{atw19}'s forest-based design with the exponential-series approach adopted by \cite{bs91} and \cite{wu10} to estimate unconditional densities. The combination of these approaches can be briefly described as follows. Let $\boldsymbol{\phi}$ be a $J$-dimensional vector of Legendre basis functions on $[0,1]$. Consider the conditional mean $\boldsymbol{\mu}_x  : = \E [ \boldsymbol{\phi} (Y)  |  X = x ] $ together with the information projection of $f(\cdot |x)$ onto the corresponding exponential family, which is defined by $\tilde{f} (y  ;  \boldsymbol{\theta}_x) : = {\exp \left[ \boldsymbol{\theta}_x^\tau  \boldsymbol{\phi} ( y ) \right]} / { \int_0^1 \exp \left[ {\boldsymbol{\theta}}_x^\tau  \boldsymbol{\phi} (t)  \right] d t }$ for $y \in [0, 1]$ and with ${\boldsymbol{\theta}}_x \in \mathbb{R}^J$ satisfying $  \int_0^1 [ \boldsymbol{\phi} (y) -   \boldsymbol{\mu}_x    ]  \exp [ {\boldsymbol{\theta}}_x^\tau  \boldsymbol{\phi} (y)  ] d y  = 0 $, i.e., matching the conditional moments. It follows from existing results that $\tilde{f}( \cdot ; \boldsymbol{\theta}_x )$ approximates $f(\cdot | x)$ as $J \rightarrow \infty$; hence, $\boldsymbol{\theta}_x$ can provide a good finite-dimensional summary about the heterogeniety of $f(\cdot | x)$ across $x$.

The proposed estimator $\hat{f}(\cdot | x)$ arises naturally from the precedent discussion and can be computed in two steps as follows. Let $\{ (Y_1, X_1) ,\dots, (Y_n , X_n) \}$ be a random sample from $(Y , X)$. In the first step, we estimate $\boldsymbol{\mu}_x$ by $\hat{ \boldsymbol{\mu}}_x = \sum_{i=1}^n \omega_i (x) \boldsymbol{\phi}  ( Y_i  )  $, where $\omega_i  (x)\in \mathbb{R} $ are weights generated by \cite{atw19}'s random forest algorithm. Specifically, each $\omega_i  (x)$ is defined as the fraction of trees in which $X_i$ appears in the same leaf as $x$. So, the weights are adaptive and generated via recursive partitioning on subsamples, where in each split we focus on maximizing the heterogeneity of $\boldsymbol{\theta}_x$ across $x$.\footnote{I refer to \cite{has09}, as well as \citet[Ch.\ 29]{ha22}, for precise definitions of machine learning concepts such as tree, leaf, and random forest.} In the second step, we set $\hat f(y|x)  =  {\exp \left[ \hat{\boldsymbol{\theta}}_x^\tau  \boldsymbol{\phi} (y) \right]} / { \int_0^1 \exp \left[ \hat{\boldsymbol{\theta}}_x^\tau  \boldsymbol{\phi} (t)  \right] d t }$, where $ \hat{\boldsymbol{\theta}}_x \in \mathbb{R}^{J}$ is obtained by solving the nonlinear system $  \int_0^1 [ \boldsymbol{\phi} (y) -   \hat{\boldsymbol{\mu}}_x    ]  \exp [ \boldsymbol{\theta}^\tau  \boldsymbol{\phi} (y)  ] d y  = 0 $ with respect to $\boldsymbol{\theta} \in \mathbb{R}^J$.

The distinguishing feature of this approach is the use of adaptive weights based on a random forest design.\footnote{Indeed, with the aim of estimating an auction model with risk-averse bidders, \cite{zin18} has suggested using kernel weights in the first step instead of $\omega_i  (x)$.} Such a choice is motivated by the increasing success of machine learning techniques and, in particular, random forest algorithms in empirical applications. Among other advantages, these algorithms have succeeded in dealing with the curse of dimensionality and screening out irrelevant covariates \cite[]{wa18}. They have also shown good predictive power in supervised learning schemes. In econometrics, machine learning techniques has been applied to a wide range of models that involve, e.g., demand estimation \cite[]{ba15machine}, as well as estimating heterogeneous treatment effects in regression discontinuity designs \cite[]{as16class}.

Regarding the asymptotic properties, I show that $\hat{f}(\cdot | x ) $ is uniformly consistent on $[ 0 , 1]$ and pointwise asymptotically normal as $J$ increases with $n \rightarrow \infty$. I also provide a ratio-consistent variance estimator to assess the precision of $\hat{f}(y | x ) $ and to build asymptotically valid confidence intervals for ${f}(y | x )$. To derive these results, first, I characterize $\hat{ \boldsymbol{\mu}}_x $ as a high-dimensional infinite-order $U$-statistic with a random kernel. Second, I adapt the arguments in \cite{bs91} to conditional densities and extend the results of \cite{wa18} and \cite{atw19} to allow $J \rightarrow \infty$, relying on inequalities from \cite{vit92} for the variance of $U$-statistics.

The proposed estimator $\hat{f}(\cdot|x)$ aims to help the applied researcher in estimating not only $f(\cdot | x)$, but also infinite-dimensional parameters in nonparemetric models such as the ones studied in \cite{mat03,mat07,mat08id,mat15}. Empirical auction models constitute another example where the proposed estimator $\hat{f}(\cdot|x)$ can be useful, as the parameters of interest can be usually expressed as a functional of the conditional density of bids. Among others, \cite{ah07} and \cite{pv21} provide comprehensive surveys on this topic. In this paper, I apply the proposed estimator  to a real-world dataset of timber auctions from the U.S.\ Forest Service and, in the Supplementary Material, I provide nonparametric estimates of the conditional density of bids given certain characteristics of the auctioned timber track.


There is an extensive body of literature on conditional density estimation. Kernel-based methods have constituted the traditional approach; however, their poor finite-sample performance under many covariates has triggered the developments of bandwidth selection methods and alternative techniques. To name a few, \cite{harali04} develop a cross-validation bandwidth selection method that is able to distinguish relevant from irrelevant covariates. \cite{efro07} studies adaptive optimal rates using orthogonal polynomials for one-dimensional continuous variables. \cite{shen2016} studies the frequentist properties of nonparametric Bayesian models, focusing on adaptive density regression in high-dimensional settings; they develop priors based on orthogonal polynomials and splines, achieving adaptive optimal contraction rates. In a similar setting, \cite{nor2017} and \cite{nor2022} use priors based on mixtures of regressions, addressing the issue of irrelevant covariates. In addition, kernel-based estimators also suffer from boundary bias problems and, in this regard, series and local polynomial techniques constitute effective solutions: see, e.g., \cite{ca22} and the references cited therein.

This paper contributes to this literature but differs from the aforementioned references in that it does not examine the effect on the convergence rate of the smoothness of $f(y |x)$ with respect to $x$. This limitation commonly arises in the study of the asymptotic properties of forest-based estimators, as in \cite{wa18} and \cite{atw19}, although there are recent notable exceptions such as \cite{mou20} and \cite{cat23inf}. Thus, this paper is closely related to recent studies that incorporate machine learning techniques into the development of conditional density estimators such as \cite{dal20} and \cite{gh22}. The distinguishing aspect of my paper, compared to these articles, is the use of \cite{atw19}'s forest design and the establishment of desired asymptotic properties. Additionally, the use of a series method ensures that the uniform consistency result applies to the entire support, preventing the proposed estimator from experiencing boundary bias problems.

The rest of this paper is organized as follows. Section \ref{sec:esti} introduces the proposed estimator together with the algorithm to compute its weights. Section \ref{sec:ap} presents the asymptotic properties and a ratio consistent estimator of the asymptotic variance. Section \ref{sec:mc} reports the results of Monte Carlo experiments. Section \ref{sec:conclu} concludes with a discussion of possible extensions. The proofs of the lemmas, theorems, and corollaries are relegated to the Supplementary Material, where I also provide additional results from Monte Carlo experiments and an empirical illustration. The following notation will be employed hereafter.

\paragraph{Notation.} An array with its elements separated by commas --such as $v =  (v_1,\dots, v_m)$\sloppy -- is always considered a column vector. The super-script $^\tau$ denotes the transpose, $\mathbf{1}_m$ is an $m$-dimensional column vector of ones, $\mathbf{I}_m$ stands for the $m \times m$ identity matrix, and $\mathbb{N}_0 = \mathbb{N} \cup \{ 0 \}$. The Euclidean and sup- norms of a real vector $v$ are denoted by $\| v \|_2$ and $\| v \|_\infty$, respectively. With a slight abuse of notation, for a matrix $w$, $\|  w \|_2$ will denote the matrix norm induced by the Euclidean norm, i.e., $\|  w \|_2 = \sup_{\| v \|_2 = 1} \| w v\|_2$. For a real-valued function $\varphi$ defined on a set $T$, the notation is $\|  \varphi \|_{T,2}  = ( \int_{T} \varphi (t)^2 dt )^{1/2}$ and $\|  \varphi \|_{T,\infty}  = \sup\{ | \varphi(t) | : t \in T  \}$. The interior of $T$ is denoted by $\mathrm{int}(T)$. For an arbitrary countable set $T$, $| T| $ stands for the number of elements. The lexicographical order is our default order relation. So, e.g., the first two elements of $\{ (4,1) , (1,2) , (3,0) , (5,6) \}$ refer to $(1,2)$ and $(3,0)$. Given an ordered set $T = \{ t_1 < t_2 < \dots \}$, we write $( v_t )_{t \in T} = ( v_{t_1}, v_{t_2},\dots )$. Whenever possible, the dependence of certain terms --such as $s$, $J$, $N$, and $\sigma$-- on the sample size $n$ will be omitted from the notation. All asymptotic results are derived as $n \rightarrow \infty$, unless otherwise stated. The symbol $\convd$ stands for convergence in distribution and w.p.a.1 abbreviates with probability approaching one. Moreover, $\mathcal{N}(0,1)$ and $z_a$ denote a standard normal distribution and its $a$-quantile, respectively. With a slight abuse of notation, $\V$ denotes either the variance of a random variable or the variance-covariance matrix of a random vector. The convention $ 0 / 0 = 0$ is adopted.


\section{The estimator}		\label{sec:esti}

Let $Z = (Y,X)$ be a continuous random vector supported on $[0,1] \times\mathscr{X}$, where $\mathscr{X} \subset \mathbb{R}^{d}$ is a compact subset and $d \in \mathbb{N}$. Assume that $\mathrm{int}( \mathscr{X}  )$ is nonempty and that the marginal p.d.f.\ of $X$ is continuous and bounded away from zero on $\mathscr{X}$. Throughout this paper, consider a fixed $x \in \mathrm{int}( \mathscr{X}  )$ and let $ f(\cdot | x)$ denote the conditional density of $Y$ given $X=x$.

The objective of this section is to build a forest-based estimator of $ f(\cdot | x)$ from a random sample $\{ Z_1  ,\dots, Z_n \}$ of $Z$, being $Z_i  = (Y_i , X_i)$. I start with an assumption that imposes smoothness conditions on $ f(\cdot | x)$, similar to those in \citet[Assumption 2]{wu10}. Let $\lowmathcal{m} \geq 2$ be an integer.

\begin{assump}	\label{ass:smooth}

$f(\cdot | \cdot )$ is strictly positive and Lipschitz continuous on $[0,1] \times \mathscr{X}$, with respect to the sup-norm, and $f(\cdot | x)$ admits $\lowmathcal{m}$ continuous derivatives on $[0,1]$.

\end{assump}

Two remarks are noteworthy. First, this assumption introduces different degrees of smoothness for $f(y | x)$ along the $y$- and $x$- directions: for a fixed $x$, we assume that $f(\cdot | x)$ has $m$ continuous derivatives, while $f(y | x) $ only needs to satisfy Lipschitz continuity when it is considered a function of both $( y , x)$. This asymmetric treatment arises because the subsequent analysis will not focus on the effect of the smoothness of $f( \cdot |  \cdot)$ on the convergence rate of the proposed estimator, nor aim to achieve the optimal convergence rate. Consequently, I impose the minimal regularity condition necessary to ensure that such an estimator remains asymptotically unbiased.\footnote{
This approach might be particularly convenient, for instance, when estimating a structural model where exogenous covariates are present in the data, but they are excluded from the theoretical analysis and the empirical content does not provide guidance on the smoothness of certain equilibrium outputs with respect to these covariates.
}

Second, the support condition $(Y,X) \in [0,1] \times \mathscr{X}$ is standard in the context of series-based estimators and it can be replaced with a more general statement of the form $(Y,X) \in\ \{ (y, x)  \in \mathbb{R}^{1 + d}   : \ushort{\lowmathcal{y}} (x)  \leq y \leq \bar{\lowmathcal{y}}(x)  , \ x \in  \mathscr{X}\}$, where $\ushort{\lowmathcal{y}} (\cdot) < \bar{\lowmathcal{y}}(\cdot)$ are continuously differentiable real-valued functions on $\mathrm{int}( \mathscr{X}  )$. In this context, if $\ushort{\lowmathcal{y}} $ and $\bar{\lowmathcal{y}}$ were known, we can work with the transformation \begin{equation*}
\left(  \frac{Y - \ushort{\lowmathcal{y}}(X)}{\bar{\lowmathcal{y}}(X) - \ushort{\lowmathcal{y}}(X)}  ,  X \right) \in [0,1] \times \mathscr{X} .
\end{equation*}
Thus, throughout this section and the next one, I work under the assumption $(Y,X) \in [0,1] \times \mathscr{X}$ to simplify the exposition. In Section \ref*{sec:ei} of the Supplementary Material, I discuss strategies for handling real-world data in situations where both $\ushort{\lowmathcal{y}} $ and $\bar{\lowmathcal{y}}$ are unknown.

Now let $\phi_{\ell}$ be an orthonormal Legendre polynomial on $[0,1]$ defined as follows: \begin{equation*}
\phi_{{\ell}} (y ) =   (-1)^\ell \sqrt{2\ell + 1 }\sum_{k^\prime=0}^\ell  \binom{\ell}{k^\prime}   \binom{\ell + k^\prime}{k^\prime} (-y)^{k^\prime}  \quad  \text{for ${\ell} \in \mathbb{N}_0$ and $y \in [0,1]$.}		
\end{equation*}For $J \in \mathbb N$, consider a density $\tilde f$ from the exponential family on $[0,1]$ of the form \begin{equation*}
\tilde f ( y ; \boldsymbol{\theta} )  =  \frac{\exp \left[ \boldsymbol{\theta}^\tau  \boldsymbol{\phi} (y) \right]}{ \int_0^1 \exp \left[  \boldsymbol{\theta}^\tau  \boldsymbol{\phi} (t)  \right] d t }     ,    
\end{equation*}where $\boldsymbol{\theta}   =  ( \theta_1 , \dots , \theta_J)  \in \mathbb R^{J}$ and $\boldsymbol{\phi} (y)  = \left( \phi_1 (y), \dots, \phi_J (y) \right)$. If we allow $J$ to grow to infinity with $n$ and if $\boldsymbol{\theta}$ is appropriately chosen, then $\tilde f ( \cdot ; \boldsymbol{\theta} ) $ can approximate $f(\cdot | x)$. Specifically, let $\boldsymbol{\theta}_x \in \mathbb R^{J} $ be the vector of the information-projection coefficients that satisfies \begin{equation}
\int \boldsymbol{\phi} (y) \tilde f \left( y  ; \boldsymbol{\theta}_x \right)  dy  = \boldsymbol{\mu}_x  : =  \E [ \boldsymbol{\phi} (Y)   | X = x ] .
\label{apeqsese}
\end{equation}The next lemma establishes that $\boldsymbol{\theta}_x$ is indeed well-defined when $J$ is sufficiently large and provides the convergence rate of $\tilde f \left( \cdot  ; \boldsymbol{\theta}_x \right)$ towards $f(\cdot | x)$ in terms of $\lowmathcal{m}$.

\begin{lemma}	\label{le:approx0}
Suppose that Assumption \ref{ass:smooth} holds. Then, for any $J$ sufficiently large, $\boldsymbol{\theta}_x$ is the unique solution of the system of nonlinear equations $\int \boldsymbol{\phi} (y) \tilde f \left( y  ; \boldsymbol{\theta} \right)  dy  = \boldsymbol{\mu}_x$ with $\boldsymbol{\theta} \in \mathbb R^{J}$. Moreover, \begin{equation}
\label{eq:approxinfo}
\left\|  \tilde f \left( \cdot  ; \boldsymbol{\theta}_x \right)   -  f (\cdot| x) \right\|_{[0,1],\infty}  = \mathcal O \left(  J^{1-\lowmathcal{m}} \right)  \  \  \text{as} \ J \rightarrow \infty    .
\end{equation}
\end{lemma}

The proof of this lemma is provided in the Supplementary Material and it follows essentially by adapting the results of \cite{bs91} to conditional densities. I remark that Legendre polynomials are employed as basis functions because of their computational convenience. However, other orthonormal basis can be used instead and, in such a case, the existence and approximation results of Lemma \ref{le:approx0} can be modified accordingly; see, e.g., \cite{bs91} and \cite{ch07}.

From Eqs.\ (\ref{apeqsese}) and (\ref{eq:approxinfo}), I propose estimating $f (\cdot| x)$ by \begin{equation*}
\hat f ( y   | x) : = \tilde f \left( y   ;  \hat{\boldsymbol{\theta}}_x \right)  \   \text{for}   \   y \in [0,1],
\end{equation*}where $\hat{\boldsymbol{\theta}}_x$ is a forest-based estimator of ${\boldsymbol{\theta}}_x $ that can be computed in two steps as follows. In the first, $\boldsymbol{\mu}_x$ is estimated by generalized random forests. In the second, letting $\hat{\boldsymbol{\mu}}_x$ be the estimator from the previous step, $\hat{\boldsymbol{\theta}}_x$ is obtained by solving the nonlinear system of equations \begin{equation*}
\int \boldsymbol{\phi} (y) \tilde f \left( y  ; \boldsymbol{\theta}  \right)  dy  = \hat{\boldsymbol{\mu}}_x  \  \text{with respect to} \  \boldsymbol{\theta} \in \mathbb R^{J} .
\end{equation*}
The rest of this section focuses on the first step as the second is computationally straightforward. In this regard, I remark that existence of $\hat{\boldsymbol{\theta}}_x$ w.p.a.1 is established in Lemma \ref{le:approx1}.3 below and that, even though there is no analytical solution for $\hat{\boldsymbol{\theta}}_x$, a numerical solution can be obtained using Newton's method: see \citet[Section 2]{wu10} for further discussion.

Consider a positive integer $s <n$, define the set $\mathscr I_{n,s} = \{ ( \iota_1, \dots, \iota_s )  \in \mathbb{N}^s :  \iota_1 < \iota_2 < \dots < \iota_s \leq n   \}$, and let $\mathscr I_{n,s}^\ast$ be a bootstrap sample from $\mathscr I_{n,s}$ of size $N$: note that $\mathscr I_{n,s}^\ast$ can be obtained by non-replacement subsampling, i.e., by drawing $N$ subsamples of size $s$ from $\{1,\dots,n\}$ without replacement.\footnote{See \citet[Section 2.2]{chk19} for further discussion on this interpretation of the resampling process.} The estimator of ${\boldsymbol{\mu}}_x$ is then defined by
\begin{equation*}
\hat{\boldsymbol{\mu}}_x =  \sum_{i = 1}^n  \omega_{i} (x) \boldsymbol{\phi}  ( Y_i  )   ,
\end{equation*}
where $\{ \omega_i (x)  \in \mathbb{R} :  \ i=1,\dots,n \}$ are similarity weights that measure the relevance of the $i$th observation to fitting $\boldsymbol{\theta}_x$, i.e.,
\begin{equation}
\label{eq:wei}
\omega_i (x) = \frac{1}{N} \sum_{\boldsymbol{\iota}^\ast \in \mathscr{I}_{n,s}^\ast}  \frac{ \mathds{1} \left[ X_i \in \mathcal{L}_x^\ast \left( Z_{\iota_1^\ast} ,   \dots,  Z_{\iota_s^\ast} \right) \right]}{ \sum_{k=1}^n  \mathds{1} \left[ X_k \in \mathcal{L}_x^\ast \left( Z_{\iota_1^\ast} ,   \dots,  Z_{\iota_s^\ast}  \right) \right]}  ,
\end{equation}
and $\mathcal{L}_x^\ast  \left( Z_{\iota_1^\ast} ,   \dots,  Z_{\iota_s^\ast} \right) $ is a subsample of $\{ X_{\iota_1} ,   \dots,  X_{\iota_s}  \}$ that contains the observations falling in the same leaf as $x$. Based on \cite{bre01}'s algorithm, a detailed procedure  for constructing $\mathcal{L}_x^\ast $ and $\omega_i (x)$ is provided in Subsection \ref{sub:weights} below, together with two suggestions for the splitting schemes.



\subsection{Choosing the weights to maximize heterogeneity}		\label{sub:weights}

This subsection provides the algorithm to compute $\mathcal{L}_x^\ast ( z_{\iota_1} ,   \dots,  z_{\iota_s} ) $ via recursive partitioning, for a given training subsample $\{  z_{\iota_1} ,   \dots,  z_{\iota_s} \}$ in which each $z_\iota := (y_\iota ,  x_{\iota} ) $ can be interpreted as a realization, or data point, of $Z_\iota = (Y_\iota , X_\iota)$. I consider two splitting schemes: one that prioritizes the heterogeneity of $\boldsymbol{\theta}_x$, which is the recommended approach in this paper, and another that targets the heterogeneity of $\boldsymbol{\mu}_x$, offering a low computational cost alternative. I remark that none of these schemes are based on large-sample optimality considerations.





For given data points $\{z_1, \dots, z_n  \}$ and $J \in \mathbb N$, every split starts with a rectangular parent node $P \subset \mathrm{int}( \mathscr{X}  ) $ containing $x$, and a subsample of indexes $\mathcal{I} = \{ \iota_1 < \iota_2 < \dots < \iota_s \}$. Then we seek to solve an optimization problem of the form
\begin{equation}
\min_{ (C_1 , C_2)}  \ \mathrm{err}(C_1 , C_2 ;   P , \mathcal{I} )    ,
\label{eq:optipro}
\end{equation}
where $ (C_1 , C_2)$ are two axis-aligned children of $P$ and $\mathrm{err}(C_1 , C_2 ;  P , \mathcal{I})$ is an error function. 
In this setting, I consider two splitting schemes whose error functions are as follows. Denote $n ( \mathcal{I} ,C) = | \{ i \in \mathcal{I} :  x_i \in   C  \} |$.

\begin{example}[Targeting heterogeneity of $\boldsymbol{\theta}_x$] 		\label{exa:theta}

Following closely \cite{atw19}, we can consider
\begin{equation*}
\mathrm{err}(C_1 , C_2 ;  P , \mathcal{I}  )  
 = 
\min_{ \mathbf{t} }  \sum_{m=1,2} \Prob \left( X \in C_m \middle| X \in P   \right) \E \left[ \left\|  \mathbf{t}    -   \boldsymbol{\theta}_X \right\|_2^2   \middle|  X \in C_m \right] ,
\end{equation*}
with the minimum taking over
\begin{equation}
\label{eq:useargmin}
\mathbf{t} \in  \underset{\boldsymbol{\theta} \in \mathbb{R}^J }{\arg\min} \left\|       \int \boldsymbol{\phi} (y) \tilde f \left( y  ; \boldsymbol{\theta} \right)  dy  -   \frac{1}{n ( \mathcal{I} , P )}\sum_{\{ i \in \mathcal{I} :  X_i \in   P \}}  \boldsymbol{\phi}  ( Y_i  )    \right\|_2^2 .
\end{equation}

\end{example}

\begin{example}[Targeting heterogeneity of $\boldsymbol{\mu}_x$]	\label{exa:mu}

We can also consider
\begin{equation*}
\mathrm{err}(C_1 , C_2 ;  P , \mathcal{I}  )  
 = 
  \sum_{m=1,2} \Prob \left( X \in C_m \middle| X \in P   \right) \E \left[ \left\|  \frac{1}{n ( \mathcal{I} ,C_m)}\sum_{\{ i \in \mathcal{I} :  X_i \in   C_m  \}}  \boldsymbol{\phi}  ( Y_i  )    -   \boldsymbol{\mu}_X  \right\|_2^2   \middle|  X \in C_m \right] .
\end{equation*}

\end{example}

Essentially, each splitting scheme involves a finite-dimensional parameter that summarizes certain aspects of the conditional distribution of $Y$ given $X=x$. In the first scheme, the error function is based on \cite{atw19}'s approach, taking $\boldsymbol{\theta}_x$ as the parameter of interest. The purpose of using the $\arg\min$ set in Eq.\ (\ref{eq:useargmin}) is to guarantee the existence of a solution to this optimization problem, in a context where $n ( \mathcal{I} ,P)$ does not necessarily grow to infinity as $n \rightarrow \infty$.
In the second scheme, we target $\boldsymbol{\mu}_x$ due to its computational simplicity: unlike the first, this approach bypasses the nonlinear optimization problem in Eq.\ (\ref{eq:useargmin}). It is worth noting that none of these functions guarantee uniqueness of the solution of the optimization problem (\ref{eq:optipro}); however, we are just seeking for one solution to this problem, which we know that exists by standard arguments. 


Solving optimization problem (\ref{eq:optipro}) is often unfeasible in practice. This becomes apparent in the splitting schemes \ref{exa:theta} and \ref{exa:mu} due to the unknown distributions of $X$ and $\boldsymbol{\theta}_X $, along with the extremely high computational cost. Hence, we consider solving an approximated version of problem (\ref{eq:optipro}). For that purpose, I introduce an approximating function $\tilde{\Delta}$ that satisfies
\begin{equation}
\mathrm{err} ( C_1 , C_2 ; P , \mathcal{I} ) 
\approx 
 - \tilde{\Delta} [ C_1 , C_2  ; \mathcal{I}  ,   \lowmathcal{t} ( P , \mathcal{I}   )  ]   ,
 \label{eq:happrox}
\end{equation}
where $\lowmathcal{t} ( P , \mathcal{I}   )$ is an estimate of the targeted parameter of the splitting scheme, computed using only the data points $\{ z_i : \ x_i \in P  ,  \  i \in \mathcal{I}    \}$. Thus, solving optimization problem (\ref{eq:optipro}) becomes approximately equivalent to maximizing $\tilde{\Delta} [ C_1 , C_2  ; \mathcal{I}  ,   \lowmathcal{t} ( P , \mathcal{I}   )  ] $, with respect to $(C_1 , C_2)$.

Before proceeding, I provide the functional forms of $\tilde{\Delta} $ and $\lowmathcal{t}$ for the Splitting Schemes \ref{exa:theta} and \ref{exa:mu}. Since the approximation result in Eq.\ (\ref{eq:happrox}) is not needed in the next section for deriving the asymptotic properties of $\hat{f} (\cdot | x)$, I just provide an heuristic discussion in this regard.\footnote{I refer to Sections 2.2 and 2.3 in \cite{atw19} for a detailed discussion and formal results related to Eq.\ (\ref{eq:happrox}), noting that such results can be applied by considering a fixed $J$: exploring the approximation error as $J \rightarrow \infty$ is beyond the scope of this paper and left for future research.}



\begin{continueexample}{exa:theta}
The suggested approximating function is
\begin{equation}
\tilde{\Delta} ( C_1 , C_2  ; \mathcal{I}  ,  \boldsymbol{\theta} )  
\ =    \
 \sum_{j=1}^J \frac{1}{n ( \mathcal{I} ,C_1)  } \left[  \sum_{ \{ i \in  \mathcal{I} : x_i \in   C_1 \} }    \rho_{i,j}  ( \boldsymbol{\theta})       \right]^2     
   + 
   \frac{1}{ n ( \mathcal{I} ,C_2)  }  \left[  \sum_{ \{ i \in  \mathcal{I} : x_i \in   C_2 \} }    \rho_{i,j}  ( \boldsymbol{\theta})       \right]^2  ,
\label{eq:multiout}
\end{equation}
where $\rho_{i,j}  \left( \boldsymbol{\theta}  \right)$ denotes the $j$th element of  $\boldsymbol{\rho}_{i}  \left( \boldsymbol{\theta}  \right) : = - \mathcal{V} ( \boldsymbol{\theta} )^{-1}  [ \int \boldsymbol{\phi} (y)  f ( y  ; {\boldsymbol{\theta}} )  dy  -  \boldsymbol{\phi}  ( y_i  )  ] $, and
\begin{equation*}
\mathcal{V} (  \boldsymbol{\theta} ) = \E_{\boldsymbol{\theta}} \left\{ \left[ \boldsymbol{\phi} (\tilde{Y})   - \E \left(\boldsymbol{\phi} (\tilde{Y})  \right)  \right   ]  \left [ \boldsymbol{\phi} (\tilde{Y})   - \E \left( \boldsymbol{\phi} (\tilde{Y}) \right)  \right ]^\tau  \right\} \  \ \text{with} \   \tilde Y \sim \tilde{f} (\cdot ;  \boldsymbol{\theta})  ;
\end{equation*}
note that $\mathcal{V} (  \boldsymbol{\theta} ) $ is symmetric and positive definite for all $\boldsymbol{\theta} \in \mathbb{R}^J$ because $\{ \phi_{1} , \dots, \phi_{J} \} $ are orthonormal; see Lemma \ref*{alemma:matrix} in the Supplementary Material. In addition, $\lowmathcal{t} ( P , \mathcal{I}   )$ is a solution of the problem
\begin{equation*}
\underset{\boldsymbol{\theta} \in \mathbb{R}^J }{\min} \  \left\|       \int \boldsymbol{\phi} (y) \tilde f \left( y  ; \boldsymbol{\theta} \right)  dy  -   \frac{1}{n ( \mathcal{I} ,P)}\sum_{\{ i \in \mathcal{I} :   x_i \in  P  \}}  \boldsymbol{\phi}  ( y_i  )    \right\|_2^2  .
\end{equation*}

Roughly speaking, here, the approximation $\mathrm{err} ( C_1 , C_2 ; P , \mathcal{I}  ) \approx \tilde{\Delta} [ C_1 , C_2  ; \mathcal{I}  ,   \lowmathcal{t} ( P , \mathcal{I}   )  ] $ follows from the following arguments. First, for given children $(C_1 , C_2)$ and a fixed $J \in \mathbb{N}$, Proposition 1 in \cite{atw19} implies that 
\begin{equation*}
 \mathrm{err}(C_1 , C_2  ;  P , \mathcal{I}  )   \ \approx  \   - {\Delta} ( C_1 , C_2 ;  P , \mathcal{I} )   + \mathcal{T} (P , \mathcal{I} ) ,
\end{equation*}
where ${\Delta} ( C_1 , C_2 ; P , \mathcal{I}) =  n ( \mathcal{I} , C_1 ) n ( \mathcal{I} , C_2)  \|  \lowmathcal{t} ( C_1 , \mathcal{I}   ) -  \lowmathcal{t} ( C_2 , \mathcal{I}   )   \|_2^2 / n ( \mathcal{I} , P)^2$ and $\mathcal{T} (P , \mathcal{I} )$ is some term that depends on certain characteristics of the parent node $P$, but not on the splitting rule, nor on $( C_1 , C_2)$.  Then, Proposition 2 in \cite{atw19} suggests that ${\Delta} ( C_1 , C_2 ;P , \mathcal{I}   )$ can be further approximated by $\tilde{\Delta} [ C_1 , C_2  ; \mathcal{I}  ,    \lowmathcal{t} ( P , \mathcal{I}   )   ] $.

\end{continueexample}

\begin{continueexample}{exa:mu}

The suggested approximating function is
\begin{equation*}
\tilde{\Delta} ( C_1 , C_2  ; \mathcal{I}  ,  \boldsymbol{\mu} )  
\ =    \
 \sum_{j=1}^J \frac{1}{n ( \mathcal{I} ,C_1)  } \left[  \sum_{ \{ i \in  \mathcal{I} : x_i \in   C_1 \} }    \rho_{i,j}  ( \boldsymbol{\mu})       \right]^2     
   + 
   \frac{1}{ n ( \mathcal{I} ,C_2)  }  \left[  \sum_{ \{ i \in  \mathcal{I} : x_i \in   C_2 \} }    \rho_{i,j}  ( \boldsymbol{\mu})       \right]^2  ,
\end{equation*}
where $\rho_{i,j}  ( \boldsymbol{\mu}) =   \phi_j  ( y_i  )  -  \mu_j$.
Moreover, we have that $\lowmathcal{t} ( P , \mathcal{I}   ) = n ( \mathcal{I} ,P)^{-1}  \sum_{\{ i \in \mathcal{I} :   x_i \in  P  \}}   \boldsymbol{\phi}  ( y_i  ) $.

Heuristically, here, the approximation result follows by applying Propositions 1 and 2 in \cite{atw19} to $J$ least-square regression frameworks, each having $\phi_1  ( Y_i  ) , \dots,  \phi_J  ( Y_i  ) $ as dependent variables. See also \cite{bre01}.

\end{continueexample}



To compute $\mathcal{L}_x^\ast ( z_{\iota_1} ,   \dots,  z_{\iota_s} ) $, next I provide a recursive partitioning algorithm that solves $\max_{C_1, C_2} \tilde{\Delta} [ C_1 , C_2  ; \mathcal{I}  ,   \lowmathcal{t} ( P , \mathcal{I}   )  ] $ in each splitting step. Before doing so, I introduce an optimization problem along with input parameters and two randomization devices. Given an integer $\ushort{k} \geq 2$, a constant $\ushort{\alpha} \in (0,1/2)$, a nonempty subset $\lowmathcal{d} \subseteq \{1,\dots,d\} $, and a parent node $P$ satisfying $| \{ i \in \mathcal{I} : x_i \in P  \} | \geq 2 \ushort{k}$, consider the optimization problem 
\begin{equation}		\label{eq:optirec}
\underset{C \subset  P}{\max}  \  \tilde{\Delta} \left( C ,  P \backslash C  ; \mathcal{I}  ,  \mathbf{t}    \right) 
\end{equation}
subject to the following constraints:
\begin{enumerate}

\item[(c1)] $x \in C $, 

\item[(c2)] $C$ is a axis-aligned children that splits $P$ through the $m$th feature of $X$ for some $m \in \lowmathcal{d}$,

\item[(c3)]  $ \min \left\{ | \{ i \in  \mathcal{I}  : i \in C  \} |  ,   |   \{ i \in  \mathcal{I}  : i \in P \backslash C \} | \right\}  \geq \max \left\{ \ushort{\alpha} | \{ i \in \mathcal{I} : x_i \in P  \}   |  , \ushort{k} \right\}$.

\end{enumerate}
Note that constraint (c3) implies that a split must put at least a fraction $\ushort{\alpha}$ of the observations of the parent node into the resulting children, $C$ and $P \backslash C$, and that both must have at least $\ushort k$ observations each: see Section 3 and Definition 4 in \cite{wa18} for further discussion on this restriction.

Consider also the following randomization devices. Let $\mathcal{W}$ be discrete random vector taking values on $\{ (w_1,\dots, w_s) \in \{ 0, 1\}^s: \ \sum_{m=1}^s w_m = \lfloor s/2 \rfloor \}$ and uniformly distributed over this set. Let $\mathcal{D}$ be a set-valued random element taking values on the power set of $\{1 ,\dots, d \}$ and having a distribution (chosen by the researcher) that satisfies $\Prob (  \mathcal{D} = \emptyset ) = 0$ and $ \min_{m = 1,\dots, d} \Prob (  \mathcal{D} = \{  m \} )  > \pi / d$ for some $\pi \in ( 0, 1)$.

The steps for computing $\mathcal{L}_x^\ast ( z_{\iota_1} ,   \dots,  z_{\iota_s} ) $ are provided in Algorithm 1.

\begin{algorithm}\small
\caption{Computing $\mathcal{L}_x^\ast ( z_{\iota_1} ,   \dots,  z_{\iota_s} ) $}
\algorithmicrequire  subsample $\mathcal{I} = \{ \iota_1 < \iota_2 < \dots < \iota_s \}$; parent node $P$, $( \ushort{k} , \ushort{\alpha} , \pi)$
\begin{algorithmic}[1]
\State ${C} \gets P$ 
\State $(w_1,\dots,w_s) \gets \text{RandomDraw}(\mathcal{W})$
\State $\mathcal{I}_0 \gets \{ \iota_m  \in \mathcal{I}  :  \ m \in \{ 1,\dots, s \}   \   \&  \  w_m = 0 \}$
\State $\mathcal{I}_1 \gets \{ i  \in \mathcal{I} \backslash \mathcal{I}_0  :  \  x_i \in  C  \} $
\While{$ | \mathcal{I}_1  | \geq 2  \ushort{k} $}
	\State $\check{\mathbf{t}} \gets \lowmathcal{t} \left( C , \mathcal{I}_1   \right) $
	\State $\lowmathcal{d} \gets \text{RandomDraw}(\mathcal{D})$ 
    \State ${C}   \gets \text{solve (\ref{eq:optirec}) setting}\  (P , \mathcal{I} , \mathbf{t}) = (C , \mathcal{I}_1 , \check{\mathbf{t}}) $
    \State $\mathcal{I}_1 \gets \{  i \in \mathcal{I}_1 : \    x_i \in C  \}$
\EndWhile
\State $\mathcal{L}_x^\ast \left( z_{\iota_1} ,   \dots,  z_{\iota_s} \right)  \gets  \{ x_i  :  i \in  \mathcal{I}_0  \   \&  \  x_i   \in  {C} \} $
\end{algorithmic}
\end{algorithm}

For given data points $\{ z_1 , \dots, z_n \}$, now the weights $\omega_i (x)$ can be computed in three steps as follows. First, draw $N$ subsamples of size $s$ from $\{1,\dots, n \}$ without replacement. Second, apply Algorithm 1 to each subsample $\{ \iota_1^\ast  <   \iota_2^\ast <  \dots <  \iota_s^\ast \} $ and obtain $\mathcal{L}_x^\ast ( z_{\iota_1^\ast} ,   \dots,  z_{\iota_s^\ast} ) $.\footnote{If the same subsample is drawn twice, or multiple times, the second step must be implemented only one time per subsample. In other words, the described procedure must be applied to each subsample, so we may not need to repeat the second step $N$ times.} Third, compute $\omega_i (x)$ by applying Eq.\ (\ref{eq:wei}) to each $i = 1,\dots,n$.\footnote{If $ x_k \notin \mathcal{L}_x^\ast ( z_{\iota_1^\ast },   \dots,  z_{\iota_s^\ast }^\ast  )  \  \forall k = 1,\dots, n$ for some subsample $\{ \iota_1^\ast  <   \iota_2^\ast <  \dots <  \iota_s^\ast \} $, then follow the convention 0/0=0.}

I highlight that Algorithm 1 does not grow the entire tree, rather it only grows the branch where $x$ resides because we focus on a fixed $x$. Such a branch is grown with a double-sample procedure as the training data is divided into two parts, $\mathcal{I}_0 $ and $\mathcal{I}_1$. The resulting branch is honest because it does not use observations in $\mathcal{I}_0 $ to make the splits in Lines 5-10 of the algorithm. In other words, the branch is grown using one sub-subsample, $\mathcal{I}_1$, while the predictions at the leaf are estimated in another. In addition, the resulting leaf comes from a random-split tree in the sense that the probability that the next split occurs along the $m$th feature of $X$ is bounded from below by $\pi /d$, regardless of the relevance of such a feature and the choice of the splitting scheme. All these properties are crucial to make Assumption \ref{ass:leafs0} in the next section feasible. 


To compute $\mathcal{L}_x^\ast \left( z_{\iota_1} ,   \dots,  z_{\iota_s} \right) $ over a grid of $x$ points, Line 8 of Algorithm 1 should be modified to store each resulting child that contains at least one grid element. Since this would increase the computational cost, we may consider using Splitting Scheme \ref{exa:mu} or a thicker grid in Line 8. However, it is worth noting that using Splitting Scheme \ref{exa:mu} still imposes the same computational burden as the one in Algorithms 1 and 2 of \cite{atw19}.

To conclude this section, I emphasize that other splitting schemes can be used in Line 8 rather than the ones presented above. In other words, we can chose different approximating function $\tilde\Delta$ for the optimization problem (\ref{eq:optirec}). So, in practice and keeping into considerations the computational resources, one can adapt the splitting schemes to target certain characteristics of the parameter of interest, which often can be written as a functional of $f(\cdot|x)$.



\section{Asymptotics}		\label{sec:ap}

In this section, I establish uniform consistency and (pointwise) asymptotic normality of $\hat f (\cdot | x)$. I also provide a computationally tractable standard error formula to build asymptotically valid confidence intervals.

As a starting point, I introduce a high-level assumption on the tuning parameters, specifically focusing on the diameter of $\mathcal{L}_x^\ast ( Z_{1} ,   \dots,  Z_{s} ) \cup \{ x\}$. This diameter, denoted as $\bar{\mathrm{d}}( \mathcal{L}_x^\ast)$, is defined as the length of the longest segment parallel to one of the axes; namely, 
\begin{equation*}
\bar{\mathrm{d}}( \mathcal{L}_x^\ast)  = \max_{m = 1,\dots,d} \mathrm{d}_m( \mathcal{L}_x^\ast)   
\end{equation*}with $\mathrm{d}_m( \mathcal{L}_x^\ast)  = \sup \left\{ |  x_m^{\prime\prime}  - x_m^{\prime} |  :  \    \{ x^{\prime\prime}  ,  x^{\prime} \} \subset \mathcal{L}_x^\ast ( Z_{1} ,   \dots,  Z_{s} ) \cup \{ x\}   \right\}$.

\begin{assump}  \label{ass:leafs0}

The following conditions hold: $s \rightarrow \infty$ as $n \rightarrow \infty$, $\mathcal{L}_x^\ast $ is built from Algorithm 1, and the tuning parameters $(\ushort{k} , \ushort{\alpha} , \pi)$, as well the initial parent node $P$, are chosen so that $\Prob \left[ \bar{\mathrm{d}}( \mathcal{L}_x^\ast)  \geq s^{-\lowmathcal{c}_b} \right] = \mathcal{O} (s^{-\lowmathcal{c}_b} ) $ for some constant $\lowmathcal{c}_b >0$.

\end{assump}

In words, this assumption requires that the terminal leaf has vanishing diameter. This will be used later to derive the asymptotic properties $\hat{f}(\cdot | x)$ since it produces a bound in the bias of $\hat{\boldsymbol{\mu}}_x$, as in Lemma 1 and Theorem 3.2 in \cite{wa18}. From these results, combined with Algorithm 1, it also follows that Assumption \ref{ass:leafs0} is automatically satisfied with
\begin{equation}		\label{eq:cbleaf}
 \lowmathcal{c}_b = \frac{\pi \log[(1-\ushort{\alpha})^{-1}] }{2 d \log(\ushort{\alpha}^{-1})}  \quad \text{for any fixed} \  ( \ushort{\alpha}  ,  \ushort{k} , \pi ) \in (0,1/5] \times \mathbb{N} \times (0,1) .
\end{equation}
To have an idea about the magnitude of $\lowmathcal{c}_b > 0$, e.g., setting $\ushort{\alpha} = 1/5$ yields $\lowmathcal{c}_b \approx 0.07 \pi /d$. Concrete suggestions for choosing these tuning parameters will be provided below in Section \ref{sec:mc} (Monte Carlo experiments).



 

Eq.\ (\ref{eq:cbleaf}) holds essentially for any error function $\mathrm{err}(\cdot , \cdot)$ and any approximating function $\tilde\Delta$. This is made possible by the randomization device introduced in Line 7 of Algorithm 1 that ensures that there will be a split in any direction with positive probability, regardless of the criterion function used in Line 8 and irrespective of the relevance of the covariate. So, under proper choices of $( \ushort{\alpha}  ,  \ushort{k} , \pi )$, Assumption \ref{ass:leafs0} remains valid even with irrelevant covariates and under practically any choice of $\tilde\Delta$. In comparison to other conditional density estimators, this flexibility comes at the price of having a bias that vanishes at a slow rate.


I also remark that the tuning parameters and the initial parent node can be chosen in any way as long as the condition $\Prob [ \bar{\mathrm{d}}( \mathcal{L}_x^\ast)  \geq s^{-\lowmathcal{c}_b} ] = \mathcal{O} (s^{-\lowmathcal{c}_b} ) $ is satisfied.
This vanishing diameter condition has been introduced as a high-level assumption for two reasons. First, to maintain generality, acknowledging that different choices of these tuning parameters might yield larger (better) values of $ \lowmathcal{c}_b $ than the one derived in Eq.\ (\ref{eq:cbleaf}). Second, to simplify the presentation of the asymptotic properties later, as the obtained rates of convergence and the additional required assumptions will be directly expressed in terms of $ \lowmathcal{c}_b $ without referring  to the tuning parameters.





Following \cite{mh16}'s approach, the next step consists in characterizing $\hat{\boldsymbol \mu}_x$ as an infinite-order $U$-statistic with a random kernel. With this aim, I introduce two random vectors indexed by the set $\mathscr{I}_{n,s}$ that will play the role of randomization devices: $(V_{\boldsymbol{\iota}})_{\boldsymbol{\iota} \in \mathscr{I}_{n,s}}$ and $(W_{\boldsymbol{\iota}})_{\boldsymbol{\iota} \in \mathscr{I}_{n,s}}$. Both are assumed to independent between each other and from $(Z_1,\dots,Z_n)$. The role of $(V_{\boldsymbol{\iota}})_{\boldsymbol{\iota} \in \mathscr{I}_{n,s}}$ is to capture the randomness in generating the bootstrap sample $\mathscr{I}_{n,s}^\ast$ or, equivalently, in drawing $N$ subsamples of size $s$ from $\{1,\dots, n\}$ without replacement. Thus, it is assumed that $(V_{\boldsymbol{\iota}})_{\boldsymbol{\iota} \in \mathscr{I}_{n,s}}$ has a multinomial distribution with $N$ trials and all cell probabilities equal to $1/\binom{n}{s}$. The second random vector, $( W_{\boldsymbol{\iota}} )_{\boldsymbol{\iota} \in \mathscr{I}_{n,s}}$, captures the randomness of $( \mathcal{W}   ,  \mathcal{D} )$ in Algorithm 1 for generating the sub-subsample and producing the splits.

Given these randomization devices, I can write \begin{equation*}
\mathcal{L}_x (   Z_{\iota_1} ,   \dots,  Z_{\iota_s} , W_{\boldsymbol{\iota}}) = \mathcal{L}_x^\ast (   Z_{\iota_1} ,   \dots,  Z_{\iota_s} )
\end{equation*}for $\boldsymbol{\iota} = ({\iota}_1, \dots, {\iota}_s) \in \mathscr I_{n,s} $ and \begin{equation*}
  \hat{\boldsymbol \mu}_x   \   = \   \frac{1}{\binom{n}{s}} \sum_{\boldsymbol{\iota} \in \mathscr{I}_{n,s}} \frac{\binom{n}{s}}{N}   V_{\boldsymbol{\iota}} \times \mathbf{H}  \left( Z_{\iota_1} ,   \dots,  Z_{\iota_s} ,    W_{\boldsymbol\iota} \right)  ,
\end{equation*}where \begin{equation*}
\mathbf{H} (z_1,\dots, z_s ,  w)      =    \frac{ \sum_{i=1}^s    \mathds{1} \left[ x_i \in \mathcal L_x  (  z_1,\dots, z_s , w  ) \right]  \boldsymbol{\phi} (y_i) }{  \sum_{k=1}^s \mathds{1} \left[ x_k \in \mathcal L_x  (z_1,\dots, z_s , w ) \right]  }.
\end{equation*}I further consider the $U$-statistic  \begin{equation*}
 \tilde{\boldsymbol \mu}_x     =  \frac{1}{\binom{n}{s}} \sum_{\boldsymbol{\iota} \in \mathscr{I}_{n,s}}  \mathbf{h}  \left( Z_{\iota_1} ,   \dots,  Z_{\iota_s}  \right)  \  \text{with}  \  \mathbf{h} ( z_{\iota_1},\dots, z_{\iota_s} )  =  \E [  \mathbf{H} (z_{\iota_1},\dots, z_{\iota_s}  W_{\boldsymbol{\iota}})    ] ,
\end{equation*}noting that $\mathbf{h} ( z_1,\dots, z_s )$ is a symmetric nonrandom kernel, and I denote by $\tilde{\boldsymbol{\theta}}_{x}$ the solution of the nonlinear system of equations $ \int \boldsymbol{\phi} (y) \tilde f ( y  ; \boldsymbol{\theta}  )  dy  = \E  ( \tilde{\boldsymbol \mu}_x )$ with respect to $\boldsymbol{\theta} \in \mathbb R^{J}$, whenever it exists. Next, I can determine the convergence rates of $  \hat{\boldsymbol \mu}_x  $, $\tilde{\boldsymbol{\theta}}_x$, and $\hat{\boldsymbol{\theta}}_x$ by introducing the next assumption that determines the rate at which the tuning parameters $(s,J,N)$ should grow towards infinity.

\begin{assump}  \label{ass:sjn}
The following conditions hold: (i) $s \asymp  n^{\beta}$ for some $ \beta \in (0,1)$, (ii) $J \asymp  n^\gamma$ for some $\gamma \in \left( 0  , \min\{ 1-\beta , 2 \beta \lowmathcal{c}_b   \} /3 \right)$, and (iii) $N$ satisfies \begin{equation*}
\binom{n}{s} \left[ N \log \binom{n}{s}   \right]^{-1}  = \lowmathcal{o} (1)  .
\end{equation*}
\end{assump}

Now I can state the first lemma.

\begin{lemma}	\label{le:approx1}
The following statements hold under Assumptions \ref{ass:smooth}-\ref{ass:sjn}.\begin{enumerate}

\item For any $c>0$, we have that $\E ( \|  \hat{\boldsymbol \mu}_x    - \tilde{\boldsymbol \mu}_x )   \|_{\infty}    )  =  \lowmathcal{o} (n^{-c})$. Moreover, \begin{equation}
\left\| \tilde{\boldsymbol \mu}_x   -  \E  \left( \tilde{\boldsymbol \mu}_x \right)   \right\|_{2}   =  \mathcal{O}_p \left(  \sqrt{\frac{J s  }{n} }  \right)  \   \ \text{and}  \  \  \left\| \E  \left( \tilde{\boldsymbol \mu}_x \right) - \boldsymbol{\mu}_x   \right\|_2   =  \mathcal{O} \left(\frac{ \sqrt{J}}{  s^{\lowmathcal{c}_b}}  \right)   .
\label{eq:lemres}
\end{equation}

\item $\tilde{\boldsymbol{\theta}}_x$ exists and is unique when $J$ is sufficiently large.

Moreover,  it satisfies $\| \tilde{\boldsymbol{\theta}}_x   -  \boldsymbol{\theta}_x   \|_2 =  \mathcal{O} ( \sqrt{J} s^{-\lowmathcal{c}_b} )  $.

\item $\hat{\boldsymbol{\theta}}_x $ exists and is unique w.p.a.1, and it satisfies $\| \hat{\boldsymbol{\theta}}_x   -  \tilde{\boldsymbol{\theta}}_x   \|_2 =  \mathcal{O}_p ( \sqrt{J s/n} )  $.

\end{enumerate}

\end{lemma}

The first part of this lemma establishes that the difference $\hat{\boldsymbol \mu}_x    - \tilde{\boldsymbol \mu}_x  $ is asymptotically negligible and converges to zero at a very fast rate, which is a consequence of Assumption \ref{ass:sjn}.(iii). This implies that $\hat{\boldsymbol{\theta}}_x$ can be treated asymptotically as the solution of the system $\int \boldsymbol{\phi} (y) \tilde f \left( y  ; \boldsymbol{\theta}  \right)  dy  =  \tilde{\boldsymbol \mu}_x $. Then, the results in Eq.\ (\ref{eq:lemres}) follows by extending the arguments in \cite{wa18} to allow $J \rightarrow \infty$. In particular, the vanishing bias result is a direct consequence of the Lipchitz continuity condition of Assumption \ref{ass:smooth} combined with the vanishing diameter condition of Assumption \ref{ass:sjn}.
Existence of both $\tilde{\boldsymbol{\theta}}_x$ and $\hat{\boldsymbol{\theta}}_x $ is obtained by adapting the arguments of Lemma 5 in \cite{bs91} to conditional densities, while their approximation rates can be obtained by standard arguments. 

As a corollary of Lemmas \ref{le:approx0} and \ref{le:approx1}, it follows that $\hat{f} ( \cdot  |  x ) $ is a uniformly consistent estimator of $f ( \cdot | x)$ on $[0,1]$.

\begin{corol}	\label{cor:unifcon}

Under Assumptions \ref{ass:smooth}-\ref{ass:sjn}, we have that \begin{equation*}
\left\|    \hat{f} ( \cdot  |  x )   -   \tilde f ( \cdot ; {\boldsymbol{\theta}}_x   )   \right\|_{[0,1] , \infty}  
=  
\mathcal{O}_p \left[ J \left( \sqrt{\frac{s}{n}} +  s^{-\lowmathcal{c}_b}   \right) \right]  
\end{equation*}and therefore $\| \hat{f} ( \cdot  |  x )   -   f ( \cdot | x)   \|_{[0,1] , \infty} =  \mathcal{O}_p  \left[ J (  \sqrt{s/n}+   s^{-\lowmathcal{c}_b}  )  + J^{1-\lowmathcal{m}}   \right] $.

\end{corol}

From this corollary, $ \hat{f} ( \cdot  |  x ) - f ( \cdot | x)$ attains the fastest uniform rate of convergence by setting $\beta = ( 1 + 2 \lowmathcal{c}_b  )^{-1}$ and $\gamma =  \lowmathcal{c}_b[ m ( 1 + 2 \lowmathcal{c}_b   ) ]^{-1}$. This leads a uniform rate of $n^{-( \lowmathcal{m} - 1) \lowmathcal{c}_b / [ m ( 1 + 2 \lowmathcal{c}_b   ) ] }$. To have an idea about its magnitude, e.g., if we set $\ushort \alpha = 1/5$ in Eq.\ (\ref{eq:cbleaf}) so that $\lowmathcal{c}_b \approx 0.07 \pi /d$, the resulting rate becomes $n^{-0.07( \lowmathcal{m}-1)/[\lowmathcal{m}(d/\pi +0.14)]}$, which indicates the optimal convergence is not achieved.


With the aim of deriving the asymptotic distribution of $\hat{f} ( \cdot  |  x )$, hereafter, I consider a fixed $ y \in [0,1]$ such that $\phi_ {{\ell}} ( y) \neq \int \phi_ {{\ell}} (t) f (t | x) dt$  for some ${\ell}  \in \mathbb{N} $.\footnote{The purpose of introducing this condition is to avoid super-consistency issues; see, e.g., Theorem 8 in \cite{wu10}.} Then, the next lemma establishes that the difference $ \hat f ( y  | x)   -   \tilde f (y ; \tilde{\boldsymbol{\theta}}_x   )  $ can be approximated by a zero-mean infinite-order $U$-statistic.

\begin{lemma}	\label{le:approx2}
If Assumptions \ref{ass:smooth}-\ref{ass:sjn} are satisfied, then \begin{equation*}
\hat f ( y  | x)   -   \tilde f \left(y ; \tilde{\boldsymbol{\theta}}_x    \right)   \   =  \      \mathcal{U}   +  \mathcal{O}_p  \left( \| \mathcal{T}_x (y) \|_2 \sqrt{J}   \| \hat{\boldsymbol{\theta}}_x   - \tilde{\boldsymbol{\theta}}_x    \|_2^2   \right)   ,
\end{equation*}where \begin{equation*}
\mathcal{T}_x (y) =   \tilde f \left (y ; \tilde{\boldsymbol{\theta}}_x    \right)  \left[  \boldsymbol{\phi} (y)  -  \int \boldsymbol{\phi} (t) \tilde f \left (t ; \tilde{\boldsymbol{\theta}}_x   \right) dt  \right]^\tau \mathcal{V} (  \tilde{\boldsymbol{\theta}}_x )^{-1}  
\end{equation*}and $\mathcal{U}$ is a zero-mean infinite-order $U$-statistic of the form  \begin{equation*}
\mathcal{U}   : =   \mathcal{T}_x (y)    \left[  \tilde{\boldsymbol \mu}_x    -  \E (\tilde {\boldsymbol \mu}_x )    \right] =   \frac{1}{\binom{n}{s}} \sum_{\boldsymbol{\iota} \in \mathscr{I}_{n,s}}  \lowmathcal{h}_y  \left( Z_{\iota_1} ,   \dots,  Z_{\iota_s}  \right)  
\end{equation*}with symmetric nonrandom kernel \begin{equation*}
\lowmathcal{h}_y  \left( z_{1} ,   \dots,  z_{s}  \right) : = \mathcal{T}_x (y)  \left\{ \mathbf{h} \left( z_{1} ,   \dots,  z_{s}  \right)   - \E  \left[ \mathbf{h}  \left( Z_{1} ,   \dots,  Z_{s}  \right) \right] \right\} .
\end{equation*}

\end{lemma}

Now consider the H\'ajek projection of $\mathcal{U}$, which is given by \begin{equation*}
\mathcal{U}^\circ =  \frac{s}{n} \sum_{i=1}^n  {g}_y   (Z_i)  \  \ \text{with} \ {g}_y   (z_1)   =    \E  \left[  \lowmathcal{h}_y  \left( z_1, Z_{2} ,   \dots,  Z_{s}  \right)    \right]  .
\end{equation*}Denote its variance by $\sigma^2 = s^2 \E [ g_y   (Z_1)^2  ] /n$, noting that its dependence on $y$ is omitted from the notation to simplify the exposition.

The next lemma is the building block for obtaining the desired asymptotic distribution.

\begin{lemma}	\label{lem:norm}
The following statements hold under Assumptions \ref{ass:smooth}-\ref{ass:sjn}.\begin{enumerate}

\item There exists a constant $c_{\ref{lem:norm}} >0 $ such that $  \E [ {g}_y   (Z_1)^2  ] \geq  c_{\ref{lem:norm}}  {      \|  \mathcal{T}_x (y)  \|_2^2} / [ s \log(s)^{d} ]	$ holds when $n$ is sufficiently large.

\item Furthermore, $\E [  {g}_y   (Z_1) ^4  ]   / \{ \E [ {g}_y   (Z_1)^2]\}^{2} = \lowmathcal{o}(n)$.

\item We have that $\E(\mathcal{U}^2) / \sigma^2 \rightarrow 1$ and therefore $[{\mathcal{U}} /\sqrt{{\E(\mathcal{U}^2)}}]   - ({\mathcal{U}^\circ}/{\sigma}) =  \lowmathcal{o}_p ( 1 )$.

\end{enumerate}

\end{lemma}

The first part of this lemma implies that \begin{equation}
\label{eq:sigmabound}
\sigma \geq \sqrt{c_3}   n^{(\beta-1)/2}   [\beta \log(n)]^{-d/2}   \| \mathcal{T}_x (y)  \|_2   ,
\end{equation}while the second establishes essentially a Lyapunov condition, from which the asymptotic distribution of $\mathcal{U}^\circ / \sigma $ can be automatically derived. The proof of these parts are mainly based on the proofs of Theorems 3.3 and 3.4 in \cite{wa18}. The key technical challenges here consist in adapting \cite{wa18}'s arguments to allow $J \rightarrow \infty$ and to deal with the fact that the Lipschitz constant of the mapping $x^\prime \mapsto \E [ \mathcal{T}_x (y) \boldsymbol{\phi} (Y) | X = x^\prime ]$ can increase with $n$. Part 3.\ of Lemma \ref{lem:norm} establishes that the distributions of $\mathcal{U}$ and $\mathcal{U}^\circ$ must be very similar. The proof of this result relies on $U$-statistic theory \cite[]{vit92} and projections \cite[Ch.\ 11]{vdVaart98}.

Now we can derive the desired asymptotic distribution. 

\begin{theorem}		\label{thm:main}
Suppose that Assumptions \ref{ass:smooth}-\ref{ass:sjn} hold, as well as $\beta > (1 + 2 \lowmathcal{c}_b)^{-1}$ and $\gamma < \beta(1 +  2 \lowmathcal{c}_b) -1$. Then,
\begin{equation}  \label{eq:asympf}
\frac{ \hat f ( y  | x)   -   \tilde f \left(y ; \tilde{\boldsymbol{\theta}}_x   \right)  }{\sigma} \ \convd  \ \mathcal{N} (0 ,1)  .
\end{equation}
As a consequence, if we also assume that $\lowmathcal{m} \geq 3$ and $\gamma >  (1- \beta )/[ 2 ( \lowmathcal{m}  - 1 )]$, we have that 
\begin{equation*}
\frac{\hat f ( y  | x)    -   f ( y  | x)   }{ \sigma }  \convd \mathcal{N} (0 ,1) .
\end{equation*}
\end{theorem}

I remark that the reason for introducing extra conditions, $\beta > (1 + 2 \lowmathcal{c}_b)^{-1}$ and $\gamma < \beta(1 + 2 \lowmathcal{c}_b) - 1$, is to eliminate the remainder $\mathcal{O}_p $-term from Lemma \ref{le:approx2}. The first condition is standard as it is also required for estimating heterogeneous treatment effects \cite[Theorem 3.1]{wa18} and parameters of interest identified via a local moment equations \cite[Theorem 5]{atw19}. Combined with Assumption \ref{ass:sjn}.(ii), it automatically implies that $(1-\beta)/3 < 2 \lowmathcal{c}_b /3 $. The second condition, $\gamma < \beta(1 + 2 \lowmathcal{c}_b) - 1$, just prevents $J$ from increasing too fast. Despite the fact that $J \rightarrow \infty$, the rate at which $\hat f ( y  | x)$ approximates $\tilde f (y ; \tilde{\boldsymbol{\theta}}_x )$ can be compared with the one obtained in Theorem 5 of \cite{atw19}; essentially, they only differ by the term $\| \mathcal{T}_x (y)  \|_2 = \mathcal{O}( \sqrt{J})$.

The discussed extra conditions also guarantee that $[ \tilde{f} ( y  ;  \tilde{\boldsymbol{\theta}}_x )   -   \tilde f ( y ; \boldsymbol{\theta}_x   )    ] / \sigma = \lowmathcal{o}(1)$. However, further requirements, namely $\lowmathcal{m} \geq 3$ and $\gamma >  (1- \beta )/[ 2 ( \lowmathcal{m}  - 1 )]$, are still needed to center the asymptotic distribution of $[ \hat f ( y  | x)    -   f ( y  | x)      ]/\sigma $ at zero by making the approximation error $[ \tilde{f} ( y  ;   \boldsymbol{\theta}_x    )   -    f ( y | x ) ]/ \sigma$ negligible.

The applicability of Theorem \ref{thm:main} stems from the ability to construct confidence intervals using the quantiles of the standard normal as critical values, provided that a valid standard error for $\hat{f}(y|x)$ is available.
Specifically, for given a $\alpha \in (0,1)$ and a valid standard error $\mathrm{se} [ \hat{f}(y|x) ] $, the confidence interval
\begin{equation*}
\mathrm{CI}_{1-\alpha}  = \left[ \hat{f}(y|x)  - z_{1-\alpha/2} \times \mathrm{se} \left[ \hat{f}(y|x) \right] ,  \hat{f}(y|x)  + z_{1-\alpha/2} \times \mathrm{se} \left[ \hat{f}(y|x) \right] \right]  ,
\end{equation*}
ensures proper asymptotic coverage.
The next corollary formalizes this statement.

\begin{corol}	\label{cor:asympcov}

Suppose that Assumptions \ref{ass:smooth}-\ref{ass:sjn} hold, as well as $\beta > (1 + 2 \lowmathcal{c}_b)^{-1}$, $\gamma < \beta(1 +  2 \lowmathcal{c}_b) -1$, $\lowmathcal{m} \geq 3$, and $\gamma >  (1- \beta )/[ 2 ( \lowmathcal{m}  - 1 )]$. Let $\mathrm{se} [ \hat{f}(y|x) ]$ be a ratio consistent estimator of $\sigma$, i.e., that satisfies $\mathrm{se} [ \hat{f}(y|x) ]   /  \sigma  \convp1$. Then, $\Prob [  f(y|x) \in \mathrm{CI}_{1-\alpha} ] \rightarrow 1-\alpha$.

\end{corol}

The next subsection provides an example of a ratio consistent estimator of $\sigma$. 

\subsection{Ratio consistent standard error formula}

This subsection provides a ratio-consistent estimator $\hat\sigma^2$ of $\sigma^2$, allowing $\hat\sigma := \sqrt{\hat\sigma^2}$ to serve as a valid standard error for $\hat{f} (y | x)$. With this estimator, we can assess the precision of $\hat{f} (y | x)$ and build asymptotically valid confidence intervals for $f (y |x)$.


Following closely \citet[Section 4]{atw19} and combining their approach with the delete-$D$ Jackknife method \cite[see, e.g.,][]{sw89}, I begin by introducing an ideal estimator of $\sigma^2$:
\begin{equation*}
\hat\sigma^2 =  \frac{n - D_{\sigma}}{D_{\sigma} } \binom{n}{D_{\sigma}}^{-1} \sum_{ \boldsymbol{\iota}  \in  \mathscr{I}_{n,n-D_{\sigma}}  } \left[    \hat{\mathcal{T}}_x (y)    \left( \tilde{\boldsymbol \mu}_{x,\boldsymbol{\iota}}    -  \tilde{\boldsymbol \mu}_x   \right)  \right]^2 ,
\end{equation*}
where $D_\sigma \in \mathbb{N}$ is a sequence satisfying $ D_\sigma  < n - s$, $\hat{\mathcal{T}}_x (y)$ is the plug-in estimator of ${\mathcal{T}}_x (y)$, i.e.,\begin{equation*}
\hat{\mathcal{T}}_x (y)     =    \tilde f \left (y ; \hat{\boldsymbol{\theta}}_x    \right)  \left[  \boldsymbol{\phi} (y)  -  \int \boldsymbol{\phi} (t) \tilde f \left (t ; \hat{\boldsymbol{\theta}}_x   \right) dt  \right]^\tau \mathcal{V} (  \hat{\boldsymbol{\theta}}_x )^{-1}  ,
\end{equation*}and $\tilde{\boldsymbol \mu}_{x,\boldsymbol{\iota}} $ denotes the estimator $\tilde{\boldsymbol \mu}_x$ computed using the subsample $\{  Z_{\iota_1},\dots,  Z_{\iota_{n-D_{\sigma}}}\}$ whose indexes are taken from $\boldsymbol{\iota} = (\iota_1 , \dots, \iota_{n-D_{\sigma}}) $, after deleting $D_\sigma$ observations from the sample. To be specific, \begin{equation*}
\tilde{\boldsymbol \mu}_{x,\boldsymbol{\iota}}  =  \binom{n - D_\sigma}{s}^{-1} \sum_{\tilde{\boldsymbol{\iota}} \in \check{\mathscr{I}}_{\boldsymbol{\iota},s} }  \mathbf{h} \left( Z_{\tilde{\iota}_1} ,   \dots,  Z_{\tilde{\iota}_s}  \right)
\end{equation*}with $\check{\mathscr{I}}_{\boldsymbol{\iota},s} = \{ ( \tilde\iota_1, \dots, \tilde\iota_s )  \in \{  \iota_1, \dots, \iota_{n-D_\sigma} \}^s :  \tilde\iota_1 < \tilde\iota_2 < \dots < \tilde\iota_s   \}$.

The idea behind $\hat\sigma^2$ is to directly approximate the variance of $\mathcal{U}$ via Jackknife methods for $U$-statistics, for which I recall that $E (  \mathcal{U}^2) \approx \sigma^2$ by Lemma \ref{lem:norm}.3, and the next theorem formalizes this intuition.


\begin{theorem}
\label{thm:se}
Suppose that Assumptions \ref{ass:smooth}-\ref{ass:sjn} hold and also $\beta > (1 + 2 \lowmathcal{c}_b)^{-1}$, $\gamma < \beta(1 +  2 \lowmathcal{c}_b) -1$, and $n/D_{\sigma} \rightarrow c_\sigma$ for some finite constant $c_\sigma >1$ as $n \rightarrow \infty$. Then,  \begin{equation*}
\frac{  \hat\sigma^2  }{  \sigma^2 } - 1 = \lowmathcal{o}_p(1).
\end{equation*}
\end{theorem}

As a consequence of Corollary \ref{cor:asympcov} and Theorem \ref{thm:se}, for a given $\alpha \in (0,1)$, we have that the confidence interval defined by
\begin{equation*}
\mathrm{CI}_{1-\alpha}  = \left[ \hat{f}(y|x)  - z_{1-\alpha/2} \times \hat\sigma ,  \hat{f}(y|x)  + z_{1-\alpha/2} \times \hat\sigma  \right]  ,
\end{equation*}
provides proper asymptotic coverage, i.e., $\Prob [  f(y|x) \in \mathrm{CI}_{1-\alpha} ] \rightarrow 1-\alpha$.


I emphasize that allowing the researcher to choose $D_\sigma$ such that $n/D_\sigma \rightarrow c_\sigma > 1$ offers more flexibility in comparison with the half-sampling method adopted in the previous literature, which essentially consists in selecting $D_\sigma = \lfloor n/2 \rfloor$ and $c_\sigma = 2$. Having an additional input parameter to choose can be helpful to improve the finite-sample performance.

Since $\hat\sigma$ is computationally unfeasible, I conclude this section by proposing a modification of the subsample scheme to make it feasible. Specifically, I suggest modifying the subsample scheme as follows:\footnote{See \cite{sl09se} for further discussion about this computational procedure.}

\begin{algorithm}\small
\caption{Modified subsample scheme to compute std.\ errors}
\algorithmicrequire  positive integers $D_\sigma < n - s$, $N > 2$, and $N_\sigma < (N-1)/2$
\begin{algorithmic}[1]
\For{$l$ in 1 to  $N_\sigma$}
	\State $\mathscr{S}_{\sigma , l}  \gets  \text{subsample of size $D_\sigma$ from} \   \{ 1 , \dots , n \}  \ \text{without replacement}$
	\State $\mathscr{S}_{2l-1} \gets \text{subsample of size $s$ from} \   \{ 1 , \dots , n \}  \backslash \mathscr{S}_{\sigma , l}  \  \text{without replacement}$ 
	\State $\mathscr{S}_{2l} \gets \text{subsample of size $s$ from} \   \{ 1 , \dots , n \}  \backslash \mathscr{S}_{\sigma , l}  \  \text{without replacement}$ 
\EndFor
\For{$l$ in $2 N_\sigma +1$ to $N$}
	\State $\mathscr{S}_{l} \gets \text{subsample from} \   \{ 1 , \dots , n \}  \ \text{without replacement}$ 
\EndFor
\end{algorithmic}
\algorithmicensure $\{\mathscr{S}_{\sigma , 1} , \dots , \mathscr{S}_{\sigma ,  N_\sigma} \}$ and $\{\mathscr{S}_{1} , \dots , \mathscr{S}_{N}\}$
\end{algorithm}

\noindent Then, the estimators $\hat{\boldsymbol\mu}_{x}$ and $\hat{f} ( \cdot | x)$ can be computed by applying the procedure described in Section \ref{sec:esti} to the subsamples $\mathscr{S}_{1} , \dots , \mathscr{S}_{N}$, while a feasible version of $\hat\sigma$ can be computed as follows:
\begin{equation*}
\hat\sigma_{\mathrm{fe}} = \sqrt{    \frac{n - D_{\sigma}}{ D_{\sigma}  N_\sigma }  \sum_{l=1}^{N_\sigma}  \left[    \hat{\mathcal{T}}_x (y)    \left( \hat{\boldsymbol \mu}_{x, - l}    -   {\boldsymbol \mu}_{x,av}  \right)  \right]^2 }   ,
\end{equation*}
where $\hat{\boldsymbol \mu}_{x, -l} $ denotes the estimator of $\boldsymbol\mu_x$, calculated solely from the subsamples $\mathscr{S}_{1} , \dots , \mathscr{S}_{N}$ that do not contain $\mathscr{S}_{\sigma ,l}$, and ${\boldsymbol \mu}_{x,av} = (1/N_\sigma) \sum_{l} \hat{\boldsymbol \mu}_{x, -l} $. Note that Lines 3-4 of Algorithm 2 ensure a minimum of two such subsamples.

\section{Monte Carlo experiments}	\label{sec:mc}

This section presents the results of Monte Carlo experiments to evaluate the finite-sample performance of the proposed estimator $\hat{f}( y | x)$ when $d = 4$. I consider the the bias, standard deviation, and the Mean Integrated Squared Error (MISE) as evaluation criteria. In addition, I analyze the performance of confidence intervals that use $\hat\sigma_{\mathrm{fe}} $ as standard error.

The design of the experiments is as follows. The vector of covariates $X \in \mathbb{R}^4$ has a multivariate normal distribution with mean $(1/2) \mathbf{1}_4$ and variance-covariance matrix $(1/8)\mathbf{I}_4$, and truncated on $[0,1]^4$. The following designs are considered for the conditional distribution of $Y$ given $X$, noting that $X_4$ is an irrelevant covariate.

\begin{itemize}		

\item[D1)] $Y|X \sim$ Beta with parameters $\alpha = 1 + X_1/4 + X_2/4$ and $\beta = 1 + X_3/2$, truncated on $[1/10,9/10]$, and rescaled so that $Y\in [0,1]$.

\item[D2)] $Y|X \sim$ Log-normal with parameters $\mu = 1/2 + X_1 + X_2$ and $\sigma^2 = 1 + (X_3-1/2)^2$, truncated on $[1/4,5]$, and rescaled so that $Y\in [0,1]$.
 
\item[D3)]  $Y|X \sim$ Gaussian mixture \begin{equation*}
 \frac{1}{2} \mathcal{N} \left( -5 -X_1 - X_2 , 18 + \frac{(X_3-1/2)^2}{10} \right) + \frac{1}{2} \mathcal{N}  \left(  5 + X_1 + X_2 , 18 + \frac{(X_3-1/2)^2}{10}   \right) ,
 \end{equation*}truncated on $[-12 , 12]$, and rescaled so that $Y\in [0,1]$.

\end{itemize}
I set $x = (1/2) \mathbf{1}_4$ and the conditional density $f(y| x )$ corresponding to each design can be visualized in Figure \ref{fig:mc} below. The design points for the values of $y$ are provided in the second column of Table \ref{table:perf} below. Two sample sizes are considered: $n =500 , 1000 $.

I have performed 500 replications for each design and sample size.\footnote{These replications were conducted using the Holland Computing Center’s Swan system at the University of Nebraska, a high-performance computing resource with 56 cores and 256GB of memory.} In each replication, first, I generated a random sample from the corresponding scenario. Then, I computed the estimator $\hat{f} ( \cdot | x)$ and the standard error $\hat\sigma_{\mathrm{fe}}$ using the subsample scheme suggested in Algorithm 2. The tuning parameters of Algorithm 1 were chosen as follows: $J=8$, $N= 2240$, $P = [1/4, 3/4]^4$, $\ushort{k}= 10$, and $\ushort{\alpha} = 0.05$. In addition, I used $s \in \{ 125 ,  250 \}$ and $s \in \{ 200, 400 \}$ as subsample sizes for $n=500$ and $n = 1000$, respectively. For choosing the covariates in each splitting step (Line 7 of Algorithm 1), I randomly chose $\min\{ \max\{ \mathrm{Poisson}(5) , 1 \} , 4\}$ integers from $\{ 1, 2, 3, 4\}$ without replacement, so $\min_{m = 1,\dots, d} \Prob (  \mathcal{D} = \{  m \} )  \approx 0.0101$ and therefore $\pi \approx 0.04$. Line 8 of Algorithm 1 was implemented using a simple grid search. The tuning parameters of Algorithm 2 were $N_\sigma = 560$ and $D_\sigma = n /20$.

I also computed two alternative estimators for comparison purposes. One of them is an oracle kernel estimator of the form $\hat{f}^{ker}( y | x) = \hat{f}_{yx}^{ker}( y , x)  / \hat{f}_x^{ker}( x) $ that correctly ignores $X_4$. As tuning parameters for this estimation, I have used a tri-weight kernel and bandwidths of the form $h_y = 1.06\hat{\sigma}_y n^{-1/8}$ and $h_{x_m} = 1.06\hat{\sigma}_{x_m} n^{-1/8} $ in the numerator and $h_{x_m} = 1.06\hat{\sigma}_{x_m} n^{-1/7}$ in the denominator for $m=1,2,3$. The other estimator is a conditional Maximum Likelihood Estimator (MLE) based on a over-fitted model that has six parameters and includes $X_4$. To estimate D1, I used the specifications $b_1 + b_2 X_1 + b_2 X_2 + b_4 X_4$ and $b_5 + b_6 X_3$ for the parameters $\alpha$ and $\beta$, respectively. To estimate D2 and D3, I employed $b_1 + b_2 X_1 + b_3 X_2 + b_4 X_4 $ and $b_5 + b_6(X_3 - 1/2)^2$ for $\mu$ and $\sigma^2$, respectively. The maximization problem was then carried out with respect to $(b_1,\dots,b_6)$ over a compact set, which included the true values of the parameters.

The results of the Monte Carlo simulations are presented in Tables \ref{table:perf}-\ref{table:inf} and Figure \ref{fig:mc} at the end of this section. Table \ref{table:perf} reports the bias and the standard deviation (in parentheses) for the design points of $y$, as well as for the MISE over $[0.15 , 0.85]$. The column `GRF' displays the results linked to the proposed estimator $\hat{f}( y| x)$, while the columns `Kernel' and `MLE' provide the results from the alternative estimators. As can be noted, the bias of $\hat{f}( y| x)$ is relatively small except at certain values of $y$ in Design D3; namely, $y = 0.250, 0.500, 0.750$. Increasing $s$ seems not to have a significant impact on the bias. The standard deviation of $\hat{f}( y| x)$ is extremely small when compared to that of the oracle kernel estimator and, as expected, it is larger than the one obtained from the MLE procedure. In terms of the MISE, the proposed estimator exhibits a very good performance and clearly dominates the oracle kernel estimator. Both the standard deviation and the MISE of $\hat{f}(  y | x)$ decreases when $n$ increases.

The performance of $\hat{f}( y | x)$ when $n = 1,000$ can be visualized in Figure \ref{fig:mc}. In this figure, the solid line represents the true conditional density for each scenario. The circles and squares correspond to the mean and median of the estimator, respectively, obtained in the simulations for each value of $y$ provided in the second column of Table \ref{table:perf}. As suggested by the symmetric limiting distribution, there is an overlap between the mean and median, and both are close to the solid line: this effect is particularly noticeable in Design D1. The black up- and down-pointing triangles are the $10^\text{th}$ and $90^\text{th}$ percentiles of $\hat{f}( \cdot| x)$, respectively. It is noteworthy that the difference between these two percentiles tends to be relatively small, as one may expect from the values of the standard deviations in Table \ref{table:perf}.

Additional experimental results are provided in the Supplementary Material. Specifically, Table \ref{table:perf} and Figure \ref{fig:mc} were replicated using subsample sizes of $s \in \{ 75,  200 \}$ and $s \in \{ 125, 350 \}$ for $n=500$ and $n = 1000$, respectively, to conduct robustness checks. The results remain largely consistent with the main findings, indicating that the conclusions are not sensitive to these variations in subsample sizes.

Finally, Table \ref{table:inf} presents the results to evaluate the performance of the 95\% confidence intervals $[ \hat{f}(y | x) \pm \hat\sigma_{\mathrm{fe}} \times z_{0.975}] $ across the design points of $y$. Given the nominal confidence level of 0.95, an ideal confidence interval should achieve coverage close to this value. Among the three designs, D1 and D2 show relatively stable coverage, with some deviations, particularly for smaller and intermediate design points. On the other hand, Design D3 exhibits more variability, with noticeably lower coverage at 0.500 that can be attributed to the bias. Table \ref{table:inf} also provides the average standard errors (shown in parentheses), which can be compared to the values in parentheses under the ‘GRF’ columns in Table \ref{table:perf}. Overall, $\hat\sigma_{\mathrm{fe}}$ effectively approximates the standard deviation of $\hat{f}(y | x)$, though it tends to overestimate it at higher values of $y$.

\begin{table}[h]
\renewcommand{\arraystretch}{1.25}
	\caption{Bias, std.\ deviation, and MISE of $\hat{f}(\cdot| x)$ from simulations.} \label{table:perf}
	\begin{center}
		\begin{scriptsize}
			\begin{tabular}{!{\vrule width 1.5pt}       l  |  l  |  l   ||   r  r |  r r    ||  r r  | r  r  !{\vrule width 1.5pt}}
				
								\thickhline
				
				\multicolumn{1}{!{\vrule width 1.5pt} l |}{\multirow{3}{*}{Design}}  &  \multicolumn{1}{l |}{\multirow{3}{*}{$y$}} & \multicolumn{1}{l ||}{\multirow{3}{*}{$f (y | x)$ }}  &  \multicolumn{4}{c||}{$n=500$} &  \multicolumn{4}{c !{\vrule width 1.5pt}}{$n=1,000$}  \\     \cline{4-11}

	\multicolumn{1}{!{\vrule width 1.5pt} c |}{} 		 & \multicolumn{1}{ c |}{}   &  &  \multicolumn{2}{c|}{GRF}    & \multicolumn{1}{c|}{\multirow{2}{*}{Kernel}} & \multicolumn{1}{c||}{\multirow{2}{*}{MLE}} & \multicolumn{2}{c|}{GRF} & \multicolumn{1}{c|}{\multirow{2}{*}{Kernel}}   & \multicolumn{1}{c!{\vrule width 1.5pt}}{\multirow{2}{*}{MLE}}   \\   \cline{4-5}  \cline{8-9}
				
				\multicolumn{1}{!{\vrule width 1.5pt} c |}{} 		 & \multicolumn{1}{ c |}{}   &  & \multicolumn{1}{ l  |}{s=125}  & \multicolumn{1}{l|}{s= 250}   & \multicolumn{1}{c|}{ } & \multicolumn{1}{c||}{ } &      \multicolumn{1}{ l  |}{s=200}  & \multicolumn{1}{l|}{s= 400}    &\multicolumn{1}{ c  |}{}    & \multicolumn{1}{c !{\vrule width 1.5pt}}{} 	 \\
				
\thickhline																					
D1	&	0.125	&	0.956	&	-0.006	&	-0.008	&	0.076	&	0	&	-0.004	&	-0.004	&	0.11	&	-0.003	\\
	&	 	&	 	&	(0.17)	&	(0.17)	&	(1.25)	&	(0.06)	&	(0.11)	&	(0.12)	&	(0.95)	&	(0.04)	\\
	&	0.25	&	1.023	&	-0.023	&	-0.022	&	0.071	&	0.003	&	-0.029	&	-0.029	&	0.084	&	-0.001	\\
	&	 	&	 	&	(0.11)	&	(0.11)	&	(1.24)	&	(0.04)	&	(0.08)	&	(0.08)	&	(1.03)	&	(0.03)	\\
	&	0.375	&	1.058	&	-0.005	&	-0.002	&	0.069	&	0.005	&	-0.021	&	-0.021	&	0.065	&	0.001	\\
	&	 	&	 	&	(0.14)	&	(0.14)	&	(1.23)	&	(0.04)	&	(0.09)	&	(0.09)	&	(1.02)	&	(0.03)	\\
	&	0.5	&	1.069	&	0.016	&	0.018	&	0.078	&	0.005	&	-0.002	&	-0.002	&	-0.01	&	0.002	\\
	&	 	&	 	&	(0.15)	&	(0.15)	&	(1.31)	&	(0.05)	&	(0.10)	&	(0.10)	&	(0.98)	&	(0.03)	\\
	&	0.625	&	1.058	&	0.021	&	0.02	&	0.039	&	0.003	&	0.013	&	0.012	&	0.06	&	0.002	\\
	&	 	&	 	&	(0.13)	&	(0.13)	&	(1.25)	&	(0.04)	&	(0.10)	&	(0.10)	&	(0.97)	&	(0.03)	\\
	&	0.75	&	1.023	&	0.006	&	0.003	&	0.158	&	-0.002	&	0.015	&	0.014	&	0.008	&	0.001	\\
	&	 	&	 	&	(0.17)	&	(0.17)	&	(1.30)	&	(0.04)	&	(0.12)	&	(0.13)	&	(0.96)	&	(0.03)	\\
	&	0.875	&	0.956	&	-0.035	&	-0.036	&	0.097	&	-0.008	&	-0.005	&	-0.005	&	0.01	&	0	\\
	&	 	&	 	&	(0.19)	&	(0.19)	&	(1.17)	&	(0.06)	&	(0.13)	&	(0.13)	&	(0.95)	&	(0.05)	\\
	&	 MISE 	&	 - 	&	\emph{0.015}	&	\emph{0.015}	&	\emph{1.136}	&	\emph{0.001}	&	\emph{0.007}	&	\emph{0.008}	&	\emph{0.681}	&	\emph{0.001}	\\
\thickhline																					
D2	&	0.125	&	1.028	&	-0.088	&	-0.083	&	0.121	&	-0.016	&	-0.084	&	-0.09	&	-0.092	&	-0.008	\\
	&	 	&	 	&	(0.16)	&	(0.16)	&	(1.37)	&	(0.10)	&	(0.12)	&	(0.12)	&	(0.91)	&	(0.07)	\\
	&	0.25	&	1.275	&	-0.156	&	-0.156	&	0.065	&	0.006	&	-0.162	&	-0.163	&	0.083	&	0.001	\\
	&	 	&	 	&	(0.13)	&	(0.13)	&	(1.35)	&	(0.07)	&	(0.10)	&	(0.10)	&	(1.21)	&	(0.05)	\\
	&	0.375	&	1.259	&	-0.012	&	-0.016	&	0.08	&	0.012	&	-0.029	&	-0.026	&	0.041	&	0.005	\\
	&	 	&	 	&	(0.17)	&	(0.17)	&	(1.49)	&	(0.06)	&	(0.12)	&	(0.13)	&	(1.17)	&	(0.04)	\\
	&	0.5	&	1.155	&	0.089	&	0.086	&	0.14	&	0.01	&	0.074	&	0.079	&	0.164	&	0.004	\\
	&	 	&	 	&	(0.15)	&	(0.16)	&	(1.39)	&	(0.04)	&	(0.12)	&	(0.12)	&	(1.09)	&	(0.03)	\\
	&	0.625	&	1.029	&	0.065	&	0.063	&	0.15	&	0.005	&	0.061	&	0.065	&	0.012	&	0.003	\\
	&	 	&	 	&	(0.13)	&	(0.13)	&	(1.28)	&	(0.04)	&	(0.10)	&	(0.10)	&	(1.01)	&	(0.03)	\\
	&	0.75	&	0.906	&	-0.032	&	-0.031	&	0.022	&	0	&	-0.024	&	-0.023	&	-0.038	&	0.001	\\
	&	 	&	 	&	(0.18)	&	(0.18)	&	(1.07)	&	(0.05)	&	(0.14)	&	(0.14)	&	(0.88)	&	(0.04)	\\
	&	0.875	&	0.794	&	-0.095	&	-0.095	&	0.022	&	-0.003	&	-0.073	&	-0.075	&	0.059	&	0	\\
	&	 	&	 	&	(0.17)	&	(0.18)	&	(1.08)	&	(0.07)	&	(0.14)	&	(0.14)	&	(0.94)	&	(0.05)	\\
	&	 MISE 	&	 - 	&	\emph{0.023}	&	\emph{0.023}	&	\emph{1.193}	&	\emph{0.002}	&	\emph{0.015}	&	\emph{0.016}	&	\emph{0.766}	&	\emph{0.001}	\\
\thickhline																					
D3	&	0.125	&	0.943	&	-0.054	&	-0.053	&	0.136	&	-0.003	&	-0.043	&	-0.042	&	0.017	&	0.001	\\
	&	 	&	 	&	(0.14)	&	(0.14)	&	(1.16)	&	(0.05)	&	(0.11)	&	(0.11)	&	(0.99)	&	(0.03)	\\
	&	0.25	&	1.208	&	-0.233	&	-0.235	&	0.039	&	0.009	&	-0.234	&	-0.236	&	0.057	&	0.005	\\
	&	 	&	 	&	(0.11)	&	(0.11)	&	(1.30)	&	(0.06)	&	(0.08)	&	(0.08)	&	(1.10)	&	(0.04)	\\
	&	0.375	&	1.094	&	-0.024	&	-0.026	&	0.084	&	0.002	&	-0.038	&	-0.041	&	0.051	&	-0.002	\\
	&	 	&	 	&	(0.13)	&	(0.13)	&	(1.30)	&	(0.04)	&	(0.09)	&	(0.09)	&	(1.08)	&	(0.03)	\\
	&	0.5	&	0.956	&	0.197	&	0.195	&	0.164	&	-0.008	&	0.176	&	0.173	&	0.124	&	-0.008	\\
	&	 	&	 	&	(0.13)	&	(0.13)	&	(1.36)	&	(0.09)	&	(0.09)	&	(0.10)	&	(0.98)	&	(0.06)	\\
	&	0.625	&	1.094	&	0.093	&	0.094	&	0.121	&	0.002	&	0.079	&	0.077	&	0.05	&	-0.002	\\
	&	 	&	 	&	(0.13)	&	(0.13)	&	(1.32)	&	(0.04)	&	(0.10)	&	(0.10)	&	(1.01)	&	(0.03)	\\
	&	0.75	&	1.208	&	-0.087	&	-0.083	&	0.146	&	0.009	&	-0.084	&	-0.083	&	0.006	&	0.005	\\
	&	 	&	 	&	(0.17)	&	(0.18)	&	(1.34)	&	(0.06)	&	(0.13)	&	(0.14)	&	(1.15)	&	(0.04)	\\
	&	0.875	&	0.943	&	-0.061	&	-0.06	&	0.011	&	-0.003	&	-0.039	&	-0.035	&	0.042	&	0.001	\\
	&	 	&	 	&	(0.18)	&	(0.18)	&	(1.17)	&	(0.05)	&	(0.12)	&	(0.13)	&	(1.00)	&	(0.03)	\\
	&	 MISE 	&	 - 	&	\emph{0.028}	&	\emph{0.028}	&	\emph{1.231}	&	\emph{0.002}	&	\emph{0.020}	&	\emph{0.021}	&	\emph{0.806}	&	\emph{0.001}	\\
\thickhline

			\end{tabular}
		\end{scriptsize}
	\end{center}

{\raggedright \footnotesize{Bias and std.\ deviation are reported in normal fonts under the columns `GRF', `Kernel,' and `MLE;' std.\ deviation is in parentheses. MISE over [0.15,0.85] is reported in \emph{italics}.}}

\end{table}

\begin{table}[h]
\renewcommand{\arraystretch}{1.25}
	\caption{Coverage probabilities and performance of std.\ error formula.} \label{table:inf}
	\begin{center}
		\begin{scriptsize}
			\begin{tabular}{!{\vrule width 1.5pt}       l  |   l   ||   r  r  |  r r   !{\vrule width 1.5pt}}
				
								\thickhline
				
				\multicolumn{1}{!{\vrule width 1.5pt} l |}{\multirow{2}{*}{Design}}  &  \multicolumn{1}{l ||}{\multirow{2}{*}{$y$}} &  \multicolumn{2}{c|}{$n=500$} &  \multicolumn{2}{c !{\vrule width 1.5pt}}{$n=1,000$}  \\     \cline{3-6}

\multicolumn{1}{!{\vrule width 1.5pt} c |}{} 		 &    \multicolumn{1}{ c ||}{}   & \multicolumn{1}{l |}{$s=125$}  & $s=250$ & $s=200$  & \multicolumn{1}{c !{\vrule width 1.5pt}}{$s=400$} \\  \cline{3-6}

	\thickhline

D1	&	0.125 	&	0.858 	&	0.846 	&	0.960 	&	0.950 	\\
	&	 	&	(0.12)	&	(0.12)	&	(0.13)	&	(0.13)	\\
	&	0.250 	&	0.996 	&	0.998 	&	1.000 	&	1.000 	\\
	&	 	&	(0.21)	&	(0.21)	&	(0.22)	&	(0.22)	\\
	&	0.375 	&	0.928 	&	0.924 	&	0.990 	&	0.990 	\\
	&	 	&	(0.13)	&	(0.13)	&	(0.14)	&	(0.14)	\\
	&	0.500 	&	0.786 	&	0.788 	&	0.930 	&	0.950 	\\
	&	 	&	(0.10)	&	(0.10)	&	(0.10)	&	(0.10)	\\
	&	0.625 	&	1.000 	&	1.000 	&	1.000 	&	1.000 	\\
	&	 	&	(0.28)	&	(0.28)	&	(0.29)	&	(0.29)	\\
	&	0.750 	&	0.704 	&	0.698 	&	0.850 	&	0.820 	\\
	&	 	&	(0.10)	&	(0.10)	&	(0.10)	&	(0.10)	\\
	&	0.875 	&	0.994 	&	0.992 	&	1.000 	&	1.000 	\\
	&	 	&	(0.30)	&	(0.30)	&	(0.31)	&	(0.31)	\\
								
								\hline
											
D2	&	0.125 	&	0.756 	&	0.758 	&	0.900 	&	0.870 	\\
	&	 	&	(0.11)	&	(0.12)	&	(0.12)	&	(0.12)	\\
	&	0.250 	&	0.952 	&	0.958 	&	0.990 	&	0.980 	\\
	&	 	&	(0.19)	&	(0.20)	&	(0.20)	&	(0.20)	\\
	&	0.375 	&	0.858 	&	0.858 	&	0.970 	&	0.970 	\\
	&	 	&	(0.12)	&	(0.12)	&	(0.13)	&	(0.13)	\\
	&	0.500 	&	0.694 	&	0.682 	&	0.850 	&	0.820 	\\
	&	 	&	(0.09)	&	(0.09)	&	(0.10)	&	(0.10)	\\
	&	0.625 	&	1.000 	&	1.000 	&	1.000 	&	1.000 	\\
	&	 	&	(0.26)	&	(0.26)	&	(0.26)	&	(0.27)	\\
	&	0.750 	&	0.638 	&	0.636 	&	0.730 	&	0.720 	\\
	&	 	&	(0.10)	&	(0.10)	&	(0.09)	&	(0.09)	\\
	&	0.875 	&	0.984 	&	0.982 	&	1.000 	&	1.000 	\\
	&	 	&	(0.27)	&	(0.27)	&	(0.27)	&	(0.27)	\\

	\hline
												
D3	&	0.125 	&	0.818 	&	0.818 	&	0.950 	&	0.950 	\\
	&	 	&	(0.11)	&	(0.12)	&	(0.12)	&	(0.12)	\\
	&	0.250 	&	0.916 	&	0.922 	&	0.990 	&	0.980 	\\
	&	 	&	(0.21)	&	(0.21)	&	(0.22)	&	(0.22)	\\
	&	0.375 	&	0.934 	&	0.936 	&	0.990 	&	0.980 	\\
	&	 	&	(0.13)	&	(0.13)	&	(0.13)	&	(0.13)	\\
	&	0.500 	&	0.500 	&	0.508 	&	0.580 	&	0.610 	\\
	&	 	&	(0.10)	&	(0.10)	&	(0.10)	&	(0.10)	\\
	&	0.625 	&	1.000 	&	1.000 	&	1.000 	&	1.000 	\\
	&	 	&	(0.30)	&	(0.31)	&	(0.31)	&	(0.31)	\\
	&	0.750 	&	0.724 	&	0.740 	&	0.800 	&	0.800 	\\
	&	 	&	(0.14)	&	(0.15)	&	(0.14)	&	(0.14)	\\
	&	0.875 	&	0.996 	&	0.996 	&	1.000 	&	1.000 	\\
	&	 	&	(0.34)	&	(0.35)	&	(0.35)	&	(0.35)	\\
											
\thickhline											

			\end{tabular}
		\end{scriptsize}
	\end{center}

{\raggedright \footnotesize{Nominal confidence level: 0.95. Coverage probabilities are shown in regular font, with the average std.\ error in parentheses.}}
	
\end{table}

\clearpage

\begin{figure}[h]

\caption{Performance of $\hat{f}(\cdot|x)$ when $n=1,000$ from Monte Carlo experiments.}

\label{fig:mc}

\begin{center}

\includegraphics[scale=.425]{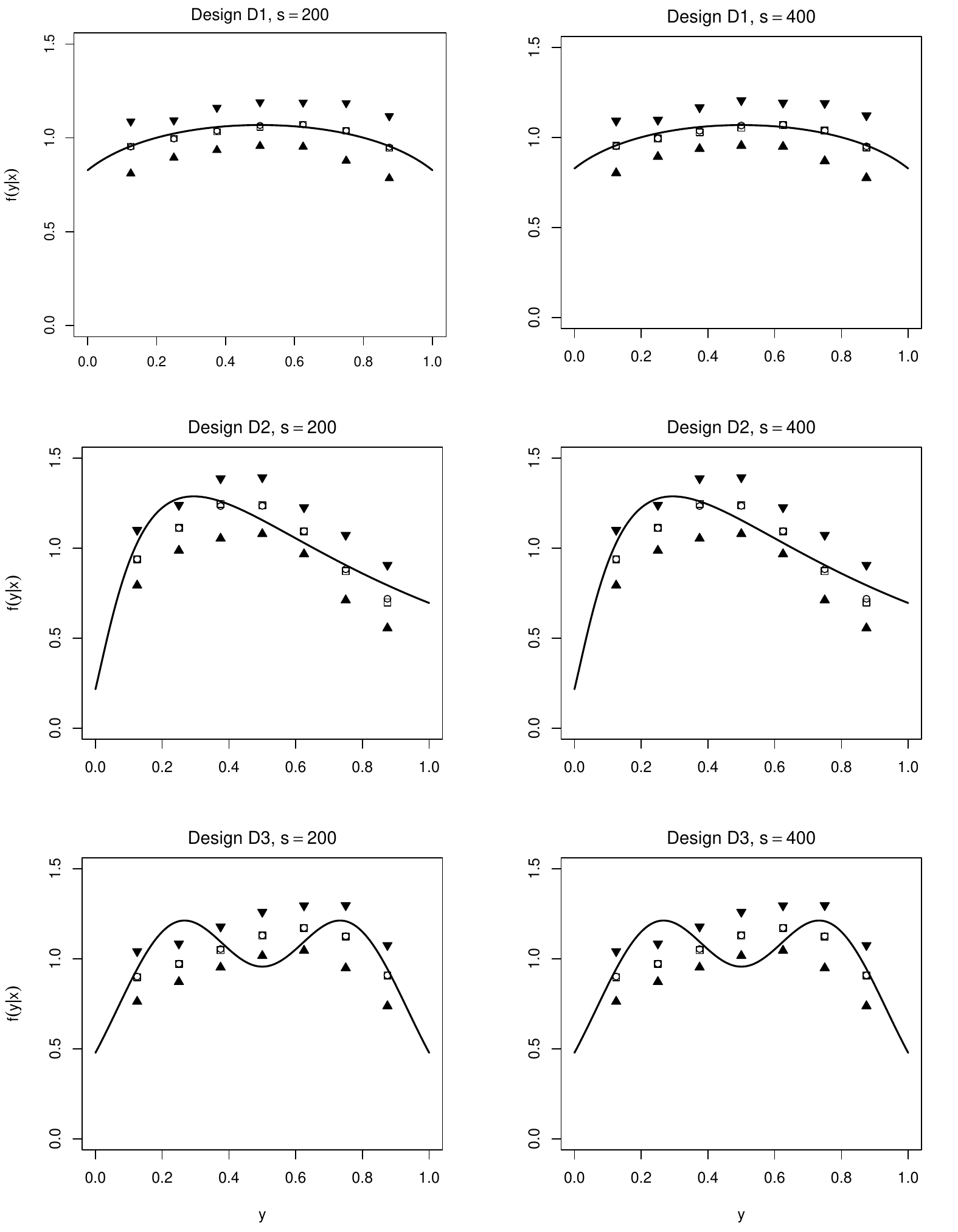}

\end{center}
{\raggedright\footnotesize{Solid line is the true conditional density with $x = (1/2) \mathbf{1}_4$. Circles and squares are the obtained mean and median of $\hat{f}( y| x)$ for the corresponding values of $y$. Black up- and down-pointing triangles are the obtained $10^\text{th}$ and $90^\text{th}$ percentiles of $\hat{f}( y| x)$, respectively.}}

\end{figure}

\section{Conclusion}		\label{sec:conclu}

I have proposed a nonparametric estimator for the conditional density of $Y$ given $X=x$. The proposed estimator $\hat{f}(\cdot |x)$ is based on \cite{atw19}'s generalized random forest design, targeting the heterogeneity of $\boldsymbol{\theta}_x$ in each splitting step. I have shown that $\hat{f}(\cdot |x)$ is uniformly consistent and asymptotically normal. I also have provided a standard error formula that allows the researcher to build asymptotically valid confidence intervals, using the percentiles of a standard normal as critical values.

I conclude with suggestions on potential avenues for future research. First, the obtained results can be generalized to scenarios where $Y$ is a random vector, i.e., $Y \in [0,1]^{d^\prime}$ with $d^\prime \in \mathbb{N}$. To do so, one can consider the multivariate orthonormal Legendre polynomials on $[0,1]^{d^\prime}$, defined by
\begin{equation*}
\phi_{\boldsymbol{\ell}} (y ) = \prod_{k=1}^{d^\prime} \phi_{\ell_k} (y_k) \  \  \text{for}  \    \boldsymbol{\ell} = ( \ell_1 ,\dots,  \ell_{d^\prime}) \in \mathbb{N}_0^{d^\prime} \  \text{and} \ y = (y_1,\dots, y_{d^\prime})  \in [0,1]^{d^\prime} ,
\end{equation*}together the following multivariate conditional density from the exponential family:
\begin{equation*}
\tilde f ( y ; \boldsymbol{\theta} )  =  \frac{\exp \left[ \boldsymbol{\theta}^\tau  \boldsymbol{\phi} (y) \right]}{ \int_{[0,1]^{d^\prime}} \exp \left[  \boldsymbol{\theta}^\tau  \boldsymbol{\phi} (t)  \right] d t }     ,    
\end{equation*}where here $\boldsymbol{\theta}   =  ( \theta_1 , \dots , \theta_J)  \in \mathbb R^{J}$, $\boldsymbol{\phi} (y)  = \left( \phi_{\boldsymbol{\ell}_1} (y), \dots, \phi_{\boldsymbol{\ell}_J} (y) \right)$, and $\{\boldsymbol{\ell}_1 < \boldsymbol{\ell}_2 < \dots < \boldsymbol{\ell}_J\}$ are the first $J$ elements of the set $ \mathbb{N}_0^{d^\prime} \backslash \{  (0,\dots,0) \}$. Then, the multivariate conditional density of $Y$ given $X=x$ can be estimated by $\hat f ( y   | x) : = \tilde f ( y   ;  \hat{\boldsymbol{\theta}}_x ) $, where $\hat{\boldsymbol{\theta}}_x$ satisfies
\begin{equation*}
\int \boldsymbol{\phi} (y) \tilde f \left( y  ; \hat{\boldsymbol{\theta}}_x  \right)  dy  =   \sum_{i = 1}^n  \omega_{i} (x) \boldsymbol{\phi}  ( Y_i  )   ,
\end{equation*}and $\{ \omega_i (x)  \in \mathbb{R} :  \ i=1,\dots,n \}$ can be obtained by adapting the arguments of Section \ref{sub:weights}. The asymptotic findings of Section \ref{sec:ap} can be extended to encompass this multivariate scenario by combining together the results of this paper with those from \cite{wu10}, which considers a multivariate unconditional density.

Second, in Section \ref{sub:weights}, the heterogeneity of $\boldsymbol{\theta}_x$ is targeted by maximizing Eq.\ (\ref{eq:multiout}) and using the pseudo-outcomes $\boldsymbol{\rho}_{i} (\boldsymbol{\theta} )$, $i=1,\dots,n$. Since $\boldsymbol{\rho}_{i} (\boldsymbol{\theta} )$ can be treated as a multivariate output, an alternative approach here might be using the methods described in \cite{sch23tree} that are based on modified splitting or stopping rules for multi-output regressions. Third, an alternative variance estimator can be constructed by combining the approximation of Lemma \ref{le:approx2} with the recent developments on variance estimation for random forests with infinite-order $U$-statistics \cite[]{xu23var}.
Fourth, it may be helpful for practical purposes to develop a selection rule for $J$, taking into account the computational challenges that this may involve. 

\clearpage

\appendix

\renewcommand{\thesection}{S.\arabic{section}}\setcounter{section}{0}

\renewcommand{\thesubsection}{S.\arabic{section}.\arabic{subsection}}\setcounter{subsection}{0}

\renewcommand{\thepage}{S.\arabic{page}}\setcounter{page}{1}

\renewcommand{\thealemma}{S.\arabic{alemma}} \setcounter{alemma}{0}

\renewcommand{\theequation}{S.\arabic{equation}} \setcounter{equation}{0}

\renewcommand{\thetable}{S.\arabic{table}}  \setcounter{table}{0}

\renewcommand{\thefigure}{S.\arabic{figure}}  \setcounter{figure}{0}

\renewcommand{\theassump}{S.\arabic{assump}} \setcounter{assump}{0}

\def\spacingset#1{\renewcommand{\baselinestretch}%
{#1}\small\normalsize} \spacingset{0}


\if0\blind
{
  \title{\bf }
  \author{ }
  \maketitle
} \fi

\if1\blind
{
  \bigskip
  \bigskip
  \bigskip
  \begin{center}
    {\HUGE\bf }
\end{center}
  \medskip
} \fi

\bigskip

  \begin{center}
    {\LARGE\bf Supplementary Material}
\end{center}

\begin{abstract}
This supplementary material is organized into three sections. The first section contains the proofs of the lemmas, theorems, and corollaries presented in the main text. The second section reports additional results from Monte Carlo experiments, and the third provides an empirical illustration
\end{abstract}

\bigskip

\spacingset{1.8} 

\section{Proofs}  \label{sec:proofs}


I start with the proof of Lemma \ref*{le:approx0}, for which I define two constants: \begin{equation*}
0 < \ushort{c}_{L\ref*{le:approx0}} : =  \min\left\{ \inf_{(y,x) \in [0 ,1] \times \mathscr{X}} f(y|x)   ,1 \right\}   \leq   \bar{c}_{L\ref*{le:approx0}} : =   \max \left\{ \sup_{(y,x) \in [0 ,1] \times \mathscr{X}} f(y|x)   ,1  \right\}  < \infty .
\end{equation*}

\paragraph{Proof of Lemma \ref*{le:approx0}.}  To establish existence of $\boldsymbol{\theta}_x$, let $(\varsigma_{\ell} )_{\ell \in \mathbb{N}_0}$ denote the Fourier coefficients of the log-density function $g (\cdot ) : =  \log[ f ( \cdot |x )]$, relative to the Legendre polynomials, and define $g_J (y) = \sum_{\ell=0}^J \varsigma_{\ell} \phi_{{\ell}} (y ) $ together with $\mathcal{T}_{L\ref*{le:approx0},1} = \| g_J  -  g \|_{[0,1],2}$ and $\mathcal{T}_{L\ref*{le:approx0},2} = \| g_J  -  g \|_{[0,1],\infty}$. Theorem 3 in \cite{bs91} guarantees that the solution of the system $\int \boldsymbol{\phi} (y) \tilde f \left( y  ; \boldsymbol{\theta} \right)  dy  = \boldsymbol{\mu}_x$, $\boldsymbol{\theta} \in \mathbb R^{J}$, exists if \begin{equation*}
4  \bar{c}_{L\ref{le:approx0}} (1/\ushort{c}_{L\ref{le:approx0}})^{1/2} \exp(4\mathcal{T}_{L\ref*{le:approx0},2} + 1   ) (J+1)  \mathcal{T}_{L\ref*{le:approx0},1}  \leq 1 .
\end{equation*}Then, the desired result is obtained by noting that $\mathcal{T}_{L\ref*{le:approx0},1} = \mathcal{O} ( J^{-\lowmathcal{m}})$ and $\mathcal{T}_{L\ref*{le:approx0},2} = \mathcal{O} ( J^{1-\lowmathcal{m}})$, which follow by Eq.\ (7.4) in \cite{bs91} and the arguments provided therein.

In view of Eq.\ (\ref{eq:approxinfo}), letting $g_x (\cdot ) : =  \log[ f ( \cdot ;  \boldsymbol{\theta}_x)]$, we can bound \begin{equation*}
\left\|  \tilde f \left( \cdot  ; \boldsymbol{\theta}_x \right)   -  f (\cdot| x) \right\|_{[0,1],\infty} \leq \exp( \bar{c}_{L\ref{le:approx0}} )  \left\|  g_x   -   g \right\|_{[0,1],\infty}  \leq  \exp( \bar{c}_{L\ref{le:approx0}} )  \left(\left\|  g_x  -   g_J \right\|_{[0,1],\infty}   +  \mathcal{T}_{L\ref*{le:approx0},2}  \right) .
\end{equation*}So, the desired result is obtained by noting that $\left\|  g_x  -   g_J \right\|_{[0,1],\infty} = \mathcal{O} ( J^{1-\lowmathcal{m}})$, which follows by adapting Lemma 5 of \cite{bs91} to conditional densities, as well as the arguments in the proof of Theorem 3 of this reference.  \qed

Before proceeding with the rest of the proofs, I introduce additional notation and an auxiliary lemma. Denote $\ushort{\boldsymbol{\iota}} = (1,2,\dots, s)$ and, for a vector $\boldsymbol{\iota} = (\iota_1, \dots, \iota_s) \in \mathscr I_{n,s}$, write $Z_{(\boldsymbol{\iota})} = (Z_{\iota_1} ,\dots, Z_{\iota_s} ) $ and $z_{(\boldsymbol{\iota})} = (z_{\iota_1} ,\dots, z_{\iota_s} ) \in ([0,1 ] \times \mathscr{X}  )^s $. Then, define \begin{equation*}
\bar{ \mathbf{H}} \left( Z_{(\boldsymbol{\iota})} , V_{\boldsymbol\iota} , W_{\boldsymbol{\iota}} \right) =\left( {\binom{n}{s}} /{N} \right)   V_{\boldsymbol\iota} \mathbf{H} (Z_{(\boldsymbol\iota)} ,  W_{\boldsymbol\iota})   -   \mathbf{h} (Z_{(\boldsymbol\iota)}) 
\end{equation*}together with $\lowmathcal{s}_i ( z_{(\boldsymbol{\iota})}  )       = \E [ \mathcal{S}_i (z_{(\boldsymbol{\iota})} , W_{\boldsymbol{\iota}} )     ]$, where \begin{equation*}
\mathcal{S}_i \left( z_{(\boldsymbol{\iota})} ,w \right)     =   \frac{ \mathds{1} \left[ x_i \in \mathcal L_x  (  z_{(\boldsymbol{\iota})} ,w  ) \right]   }{\sum_{k \in \{ \iota_1, \dots, \iota_s\}} \mathds{1} \left[ x_k \in \mathcal L_x  ( z_{(\boldsymbol{\iota})} , w ) \right]  }  \  \ \text{for} \  i \in \{ \iota_1, \dots, \iota_s\}   \ \text{and}  \   w\in \{ 0, 1\}^s .
\end{equation*}Define also the mapping $\psi : \mathbb R^{J} \rightarrow \mathbb R$ by \begin{eqnarray*}
\psi (\boldsymbol{\theta})   & : = &  \left[  \boldsymbol{\phi} (y)  -  \int \boldsymbol{\phi} (t) \tilde f \left (t ; \tilde{\boldsymbol{\theta}}_x   \right) dt  \right]^\tau  \mathcal{V} (  \tilde{\boldsymbol{\theta}}_x )^{-1}   \int  \boldsymbol{\phi} (t)  \tilde  f \left( t ; \boldsymbol{\theta} \right) d t  \\
& = & \tilde{f} (y ; \tilde{\boldsymbol{\theta}}_x )^{-1}  \int \mathcal{T}_x (y) \boldsymbol{\phi} (t)  \tilde  f \left( t ; \boldsymbol{\theta} \right) d t   ,
\end{eqnarray*}as well as the function $\tilde{\mathcal{T}} (y ; \boldsymbol{\theta}  ) = \tilde f \left (y ; {\boldsymbol{\theta}}   \right)  \left[  \boldsymbol{\phi} (y)  -  \int \boldsymbol{\phi} (t) \tilde f \left (t ; {\boldsymbol{\theta}}   \right) dt  \right]^\tau \mathcal{V} (  \boldsymbol{\theta} )^{-1}$ and the set  \begin{equation*}
 \Theta_n = \left\{   \boldsymbol{\theta} \in \mathbb{R}^J : \left\|  \boldsymbol{\theta}  -  \boldsymbol{\theta}_x   \right\|_2    < ( \sqrt{J s/n} + \sqrt{J} s^{-\lowmathcal{c}_b} )^{2/3}   \right\} \ \ \text{for}  \  n \in \mathbb{N} .
\end{equation*}Note that $\tilde{\mathcal{T}} (y ; \tilde{\boldsymbol{\theta}}_x  ) = \mathcal{T}_x (y )$. Observe also that, under Assumptions \ref{ass:smooth} and \ref{ass:sjn}, we can find $( \ushort{f} , \bar{f} , \bar{n}) \in \mathbb{R}^2 \times \mathbb{N}$ such that \begin{equation*}
0< \ushort{f} < \ushort{c}_{L\ref{le:approx0}}   \leq   \bar{c}_{L\ref{le:approx0}}   <  \bar{f} < \infty
\end{equation*}and $\ushort{f} < f( y ; \boldsymbol{\theta}  )<  \bar{f}$ for all $(y , \boldsymbol{\theta}) \in [ 0, 1]  \times \Theta_n $ and $n \geq \bar{n}$: this result follows from Lemma \ref{le:approx0}, \citet[Lemma 4, Eq.\ (5.2)]{bs91}, and  Eq.\ (\ref{eq:approxinfo}). Furthermore, let $\mathcal H_{f} ( y ; {\boldsymbol{\theta}} )$ and $\mathcal{H}_{\psi}  ( \boldsymbol{\theta} ) $ denote the Hessian matrices at $\boldsymbol{\theta}$ of $\tilde f (y ; \cdot )$ and $\psi (\cdot)$, respectively.

Now we are ready to state the auxiliary lemma, with its proof available in Subsection \ref{sub:palemma} at the end of this appendix.

\begin{alemma}		\label{alemma:matrix}

If Assumptions \ref{ass:smooth} and \ref{ass:sjn} are satisfied, then the next inequalities hold for every $n \geq \bar{n}$ and $( v , \boldsymbol{\theta} , \boldsymbol{\theta}^\prime)  \in \mathbb{R}^J  \times \Theta_n  \times \Theta_n$.
\begin{eqnarray*}
(a)  &  &   \ushort{f} \| v \|_2 \leq v^\tau \mathcal{V} ( \boldsymbol{\theta}  ) v \leq \bar{f} \| v \|_2^2 ,  \  \|  \mathcal{V} ( \boldsymbol{\theta}  ) \|_2 \leq  \bar{f} ,  \   \|  \mathcal{V} ( \boldsymbol{\theta}  )^{-1} \|_2 \leq 1/ \ushort{f},  \ \text{and}   \\
& &   \quad  \quad   \| \tilde{\mathcal{T}} (y ; \boldsymbol{\theta}  )  \|_2 \leq 2 (  \bar{f}  / \ushort{f})\sqrt{J}  . \\
(b)  &  &  | v^\tau \mathcal H_{f}   \left( y ;  \boldsymbol{\theta} \right)    v  |  \leq (\bar{f}/\ushort{f}  )^3  (  \|  \tilde{\mathcal{T}} (y ; \boldsymbol{\theta} )\|_2^2 + 1  )  \|  v \|_2^2  \   \text{and} \\
&  & \quad   \quad   | v^\tau \mathcal H_{\psi}  ( \boldsymbol{\theta} )  v | \leq 2 \sqrt{J} ( \bar{f} / \ushort{f})   \|  \tilde{\mathcal{T}} (y ; \boldsymbol{\theta} )\|_2 \| v \|_2^2  . \\
(c)   &  &  \|   \tilde{\mathcal{T}} (y ; \boldsymbol{\theta} )   -   \tilde{\mathcal{T}} (y ; \boldsymbol{\theta}^\prime ) \|_2   \leq  20(\bar{f}/\ushort{f}  )^2  J \| \boldsymbol{\theta}  -   \boldsymbol{\theta}^\prime \|_2 . 
\end{eqnarray*}

\end{alemma}

Now we are ready to state the proofs of Lemma \ref{le:approx1} and the other results outlined in the main text.

\paragraph{Proof of Lemma \ref{le:approx1}.} To prove part 1., I observe that the arguments in Step 1 in the proof of Theorem 2.1 in \citet[pp.\ 4822-4823]{song19} imply that there exists a constant $c_1 >0$ such that \begin{equation*}
\E  \left[  \max_{j} \left| \sum_{\boldsymbol\iota \in \mathscr I_{n,s}} \bar{ H_j} ( z , V_{\boldsymbol{\iota}} ,  W_{\boldsymbol\iota} ) \right| \right]  \leq c_1 \sqrt{\log(J) \binom{n}{s} \max_j  \mathfrak{B}_{j} (z) ^2   }  + s \log(J)  \log( n J )   \max_j \mathfrak{B}_{j} (z)    
\end{equation*}for any $z \in [0,1 ] \times \mathscr{X}$, where the maximum is taken over $j = 1,\dots, J$ and \begin{equation}	\label{eq:le1}
\mathfrak{B}_{j} (z)  : = \E [ \exp(| \bar{ H_j} ( z ,V_{\boldsymbol\iota},   W_{\boldsymbol\iota} )    | ) - 1 ]  \leq \exp(1) \left\{ \binom{n}{s}^{-1} \exp\left[  \binom{n}{s}   / N\right] \right\}^N .
\end{equation}I remark that the inequality in Eq.\ (\ref{eq:le1}) can be derived from the law of iterated expectations and the form of the moment generating function of a multinomial distribution. Then, as $\hat{\mu}_{x,j}    - \tilde{\mu}_{x,j}  = \binom{n}{s}^{-1}  \sum_{\boldsymbol\iota} \bar{ H_j} ( z , V_{\boldsymbol{\iota}} ,  W_{\boldsymbol\iota} ) $, $\E ( \|  \hat{\boldsymbol \mu}_x    - \tilde{\boldsymbol \mu}_x   \|_{\infty}    )  =  \lowmathcal{o} ( n^{-c}) $ follows by Assumption \ref{ass:sjn}.(iii), for any fixed $c >0$.


To derive the second result, for any $c_2 > 0$, note that we can bound $\Prob ( \| \tilde{\boldsymbol \mu}_x  - \E ( \tilde{\boldsymbol \mu}_x  ) \|_2^2 \geq c_2 s J /n  ) \leq  \max_j \E \{ [ \tilde{\mu}_{x,j}  - E (\tilde{\mu}_{x,j}) ]^2 \} n /(c_2 s)$ by using Markov's inequality. The desired result then follows from Theorem 5.2 in \cite{hoef48}, which implies $\V ( \tilde{\mu}_{x,j}  )  \leq (s/n) \V[ h_j (Z_{(\ushort{\boldsymbol\iota})})] \leq s/n$ for all $j \in \mathbb N$.

To derive the third result, which is related to the bias, observe that we can write \begin{equation}		\label{eq:reph}
\mathbf{H} (z_{(\boldsymbol{\iota})}  ,w)  = \sum_{i \in \{ \iota_1, \dots, \iota_s \}} \mathcal{S}_i ( z_{(\boldsymbol{\iota})} , w)  \boldsymbol{\phi} (y_i) \   \text{ and }  \   \mathbf{h} (Z_{(\boldsymbol{\iota})} )  = \sum_{i \in \{ \iota_1, \dots, \iota_s \}}  \E \left[ \mathcal{S}_i ( Z_{(\boldsymbol{\iota})} , W_{\boldsymbol{\iota}} )  \boldsymbol{\phi} (Y_i) | Z_i  \right]  .
\end{equation}As a result, $\E \left( \tilde{\boldsymbol\mu}_{x}   - \boldsymbol{\mu}_{x}   \right) = \E \left\{  \sum_{i=1}^s \mathcal{S}_i ( Z_{(\ushort{\boldsymbol{\iota}})} , W_{\ushort{\boldsymbol{\iota}}} )  [  \boldsymbol{\phi} (Y_i)  -  \boldsymbol{\mu}_{x}] \right\}$ follows by the law of iterated expectations. Then, the desired result follows from Assumptions  \ref{ass:smooth}-\ref{ass:leafs0} and by arguments similar to the ones in \citet[Proof of Theorem 3.2]{wa18}; specifically, for every $j \in \mathbb{N}$, we can bound \begin{multline*}
\left| \E \left( \tilde{\mu}_{x,j}   - \mu_{x,j}   \right)  \right|  = \left|  \E \left\{ \left[ \phi_j (Y_1)  - \mu_{x,j}  \right] |  X_1 \in \mathcal{L}_x (  Z_{(\ushort{\boldsymbol\iota})} ,  W_{\ushort{\boldsymbol\iota}} )   \right\}  \right|  \\
\ \   \leq \  2 \Prob\left\{ \bar{\mathrm{d}} \left[ \mathcal{L}_x ( Z_{(\ushort{\boldsymbol\iota})} ,  W_{\ushort{\boldsymbol\iota}}) \right] > s^{-\lowmathcal{c}_b} \right\}  +   \sup_{ x_1}  \int | \phi_j (t)|  | f(t |x_1)  - f (t | x) | dt ,
\end{multline*}where the supremum is taken over $\{ x_1  \in  \mathscr{X} :   \| x_1 -  x\|_\infty \leq  s^{-\lowmathcal{c}_b} \}$.

Considering now part 2., existence and uniqueness of $\tilde{\boldsymbol{\theta}}_x$ for sufficiently large $J$ follows
by Lemma \ref{le:approx0}, Assumption \ref{ass:sjn}, and by extending the arguments in the proof of Lemma 5 in \cite{bs91} to conditional densities; see also Lemma A.3 in \cite{wu10}. The convergence rate of $\|  \tilde{\boldsymbol \theta}_x  - \boldsymbol{\theta}_x  \|_2 $ can then be obtained by Eq.\ (5.6) in \cite{bs91}. Part 3.\ follows by the same arguments. \qed

\paragraph{Proof of Corollary \ref{cor:unifcon}.}  Note that \begin{equation*}
\frac{\partial \tilde{f}( y ; \boldsymbol{\theta} ) }{\partial \boldsymbol{\theta}}    =  \left[ \boldsymbol{\phi} (y) -  \int\boldsymbol{\phi} (t) \tilde{f}( t ; \boldsymbol{\theta} )  dt \right]\tilde{f}( y ; \boldsymbol{\theta} )  \  \  \text{for} \ \boldsymbol{\theta} \in \mathbb{R}^J ,
\end{equation*}so the mean value theorem and Cauchy-Schwarz inequality imply \begin{equation*}
\left| \tilde{f}( y ; \hat{\boldsymbol{\theta}}_x )   -  \tilde{f}( y ; \boldsymbol{\theta}_x )  \right| \leq 2 \bar{f} \sqrt{J}  \left\|   \hat{\boldsymbol{\theta}}_x   - \boldsymbol{\theta}_x   \right\|_{2}
\end{equation*}whenever $\hat{\boldsymbol{\theta}}_x  \in \Theta_n$, which occurs w.p.a.1 due to Lemma \ref{le:approx1} (parts 2 and 3). Then, the desired results follow immediately by combining this inequality with Lemmas 1 and 2; in particular, the uniform rate of convergence of $\hat{f} ( \cdot  |  x ) $ towards $f ( \cdot  |  x ) $ can be easily obtained from Lemma \ref{le:approx0} and the triangle inequality. \qed

\paragraph{Proof of Lemma \ref{le:approx2}.} As a starting point, I observe that the equality \begin{multline} \label{eq:approxtay}
 \tilde f \left(y ; \hat{\boldsymbol{\theta}}_x   \right)   -   \tilde f \left(y ; \tilde{\boldsymbol{\theta}}_x   \right)            =      \mathcal{T}_x (y)   \left[ \hat{\boldsymbol \mu}_x  - \E(\tilde{\boldsymbol \mu}_x ) \right]      \\
  +   \frac{1}{2}   \left(   \hat{\boldsymbol{\theta}}_x   -   \tilde{\boldsymbol{\theta}}_x   \right)^\tau \mathcal H_{f} \left( y ; \check{\boldsymbol{\theta}} \right)  \left(   \hat{\boldsymbol{\theta}}_x   -   \tilde{\boldsymbol{\theta}}_x  \right)   -    \frac{ \tilde f \left(y ; {\boldsymbol{\theta}}_x   \right)}{2} \left(\hat{\boldsymbol{\theta}}_x -  \tilde{\boldsymbol{\theta}}_x  \right)^\tau  \mathcal{H}_{\psi} (\check{\boldsymbol{\theta}}^\prime )   \left(\hat{\boldsymbol{\theta}}_x - \tilde{\boldsymbol{\theta}}_x  \right)   
\end{multline}holds w.p.a.1, for some $\check{\boldsymbol\theta}$ and $\check{\boldsymbol\theta}^{\prime}$ the segment between $\hat{\boldsymbol\theta}_x$ and $\tilde{\boldsymbol\theta}_x$. This result can be derived in three steps as follows. First, a second-order Taylor expansion yields \begin{eqnarray} 
&  & \tilde f \left(y ; \hat{\boldsymbol{\theta}}_x   \right)     -    \tilde f \left(y ; \tilde{\boldsymbol{\theta}}_x   \right)  \label{eq:approxtay1}  \\
&   & \quad  =  \   \tilde f \left (y ; \tilde{\boldsymbol{\theta}}_x   \right)   \left[  \boldsymbol{\phi} (y)  -  \int \boldsymbol{\phi} (t) \tilde f \left (t ; \tilde{\boldsymbol{\theta}}_x   \right) dt  \right]^\tau    \left(   \hat{\boldsymbol{\theta}}_x   -   \tilde{\boldsymbol{\theta}}_x   \right)  \nonumber  \\
&  &  \quad  \quad  \quad + \   \frac{1}{2} \left(   \hat{\boldsymbol{\theta}}_x   -   \tilde{\boldsymbol{\theta}}_x  \right)^\tau \mathcal H_{f} \left( y ; \check{\boldsymbol{\theta}} \right)  \left(   \hat{\boldsymbol{\theta}}_x   -   \tilde{\boldsymbol{\theta}}_x   \right)  , \nonumber
\end{eqnarray}for some $\check{\boldsymbol{\theta}} $ in the segment between $\hat{\boldsymbol{\theta}}_x $ and $\tilde{\boldsymbol{\theta}}_x $. Second, by construction of $\psi$ and since the equality $ \hat{\boldsymbol \mu}_x   -  \E(\tilde{\boldsymbol \mu}_x ) = \int \boldsymbol{\phi} (y)  \left[ \tilde  f \left( y ; \hat{\boldsymbol{\theta}}_x \right)   -  \tilde f \left( y ; \tilde{\boldsymbol{\theta}}_x \right) \right] dy$ holds w.p.a.1 (Lemma \ref{le:approx1}), we have that \begin{multline}	\label{eq:approxtay2}
 \left[  \boldsymbol{\phi} (y)  -  \int \boldsymbol{\phi} (t) \tilde f \left (t ; \tilde{\boldsymbol{\theta}}_x   \right) dt  \right]^\tau  \mathcal{V} (  \tilde{\boldsymbol{\theta}}_x )^{-1} \left[ \hat{\boldsymbol \mu}_x  - \E(\tilde{\boldsymbol \mu}_x ) \right]   \ = \ \psi \left(\hat{\boldsymbol\theta}_x \right)    -   \psi \left( \tilde{\boldsymbol{\theta}}_x \right)  \\ 
= \   \left[  \boldsymbol{\phi} (y)  -  \int \boldsymbol{\phi} (t) \tilde f \left (t ; \tilde{\boldsymbol{\theta}}_x   \right) dt  \right]^\tau    \left(   \hat{\boldsymbol{\theta}}_x   -   \tilde{\boldsymbol{\theta}}_x   \right)  +  \frac{ 1}{2} \left(\hat{\boldsymbol{\theta}}_x -\tilde{\boldsymbol{\theta}}_x \right)^\tau  \mathcal{H}_{\psi} (\check{\boldsymbol{\theta}}^\prime )   \left(\hat{\boldsymbol{\theta}}_x -\tilde{\boldsymbol{\theta}}_x \right) ,
\end{multline}where the second line arises from a second-order Taylor expansion on $\psi $ about $\tilde{\boldsymbol{\theta}}_x $. Third, Eq.\ (\ref{eq:approxtay}) follows immediately from Eqs.\ (\ref{eq:approxtay1}) and (\ref{eq:approxtay2}). Finally, the desired result can be obtained by combining Eq.\ (\ref{eq:approxtay}) with Lemmas \ref{alemma:matrix} and \ref{le:approx1}. \qed

\paragraph{Proof of Lemma \ref{lem:norm}.} Throughout this proof, I abbreviate $\mathcal{S}_i = \mathcal{S}_i (Z_{\ushort{\boldsymbol\iota}} , W_{\ushort{\boldsymbol\iota}} ) $ and $\tilde{Y}_i =  \mathcal{T}_x (y)   \boldsymbol{\phi} ( Y_i )$. I begin by stating the following result: there is a constant $c_3^{\prime} >0 $ such that \begin{equation}
\E \left\{ \left[{ g_y ( Z_1   )}/ ({ \sqrt{J}  \|\mathcal{T}_x (y)\|_2 })  \right]^2 \right\} \geq  c_3^{\prime}     \V \left[ \tilde{Y}_1 / (\sqrt{J} \|\mathcal{T}_x (y)\|_2 ) \middle| X_1 = x  \right]   \V \left[ \E ( \mathcal{S}_1     |  Z_1 )  \right]
\label{eq:boundvar}
\end{equation}for $n$ sufficiently large. Eq.\ (\ref{eq:boundvar}) can be obtained by following closely the arguments in the proof of Theorem 3.3 in \cite{wa18}; specifically, Eq.\ (37) in its supplemental material. Such arguments can be applied here because the conditional expectations $\E [ \tilde{Y}_1  / (\sqrt{J} \|\mathcal{T}_x (y)\|_2 )  | X_1 = x^\prime ]$ and $\E \{ [  \tilde{Y}_1  / (\sqrt{J} \|\mathcal{T}_x (y)\|_2 )    ]^2    | X_1 = x^\prime \}$ are both Lipschitz continuous in $x^\prime$ on $\mathrm{int}(\mathscr{X})$, and uniformly across $n$, due to Assumption \ref{ass:smooth}.


Part 1.\ can now be obtained by combining Eq.\ (\ref{eq:boundvar}) with the following two observations. First, observe that $\V \left[  \mathcal{T}_x (y)   \boldsymbol{\phi} ( Y ) | X = x  \right] \geq   \ushort{f}  \|\mathcal{T}_x (y)\|_2^2$ because all the eigenvalues of $\V [\boldsymbol{\phi} ( Y ) | X = x]$ must be greater than or equal to $\ushort{f} > 0$ and the elements of $\boldsymbol{\phi}$ are orthonormal. Second, by \citet[Lemma 3.2]{wa18}, we have $s \V [ \E ( \mathcal{S}_1   |  Z_1 ) ] \geq c_3^{\prime\prime}  /  \log(s)^{d} $ for $n$ sufficiently large and for some constant $c_3^{\prime\prime} >0$ that depends on $\ushort{k}$, $d$, and the p.d.f.\ of $X$.

Part 2.\ can be derived by following closely the arguments in the proof of Theorem 3.4 in \cite{wa18}. For the sake of completeness, the main arguments are presented in the rest of this paragraph. By honest tree ($\E (\mathcal{S}_i | Z_i)  = \E (\mathcal{S}_i | X_i) $), Jensen's inequality, and the representations in Eq.\ (\ref{eq:reph}), we can bound\begin{multline}	\label{eq:g4}
8^{-1}\E [ g_y   (Z_1)^4  ]  \ \leq  \  \E \left\{\left| \E \left[ \mathcal{S}_1 \left( \tilde{Y}_1  -  E ( \tilde{Y}_1 | X_1)  \right)  | Z_1 \right] \right|^4     \right\}  \\ 
+ \  \    \E \left\{\left| \E  \left[ \sum_{i=1}^s  \mathcal{S}_i \E (  \tilde{Y}_i  | X_i ) \middle|  Z_1 \right]   -  \E   \left( \sum_{i=1}^s  \mathcal{S}_i \tilde{Y}_i  \right)  \right|^4    \right\} ,
\end{multline}as well as $ \E \left\{\left| E \left[ \mathcal{S}_1 \left( \tilde{Y}_1  -  E ( \tilde{Y}_1 | X_1)  \right)  | Z_1 \right] \right|^4     \right\} \leq 2 \| \mathcal{T}_x (y)  \|_2^4  J^2 \E \left[ \E ( \mathcal{S}_1    |  Z_1 )^2  \right]$ and  \begin{equation*}
  \E \left\{\left| \E  \left[ \sum_{i=1}^s  \mathcal{S}_i \E (  \tilde{Y}_i  | X_i ) \middle|  Z_1 \right]   -  \E   \left( \sum_{i=1}^s  \mathcal{S}_i \tilde{Y}_i  \right)  \right|^4    \right\}  \leq 4 \| \mathcal{T}_x (y)  \|_2^4  J^2 \E \left[ \E ( \mathcal{S}_1    |  Z_1 )^2  \right] 
\end{equation*}due to the fact that $\sup_{x^\prime }\E ( | \tilde{Y}_1 | | X_1 =  x^\prime  )  \leq \| \mathcal{T}_x (y)  \|_2 \sqrt{J} $, which follows by Cauchy-Schwarz inequality. Then, noting that $\E (\mathcal{S}_1) = \E [\sum_{i=1}^s\mathcal{S}_i] /s= s^{-1}$, so $ \V \left[ \E ( \mathcal{S}_1     |  Z_1 )  \right]  =   \E \left[ \E ( \mathcal{S}_1    |  Z_1 )^2  \right]  - s^{-2}$, and combining Eqs.\ (\ref{eq:boundvar}) and (\ref{eq:g4}) yield the inequality \begin{equation*}
\frac{\E [  {g}_y   (Z_1) ^4  ] }{n  \{ \E [ {g}_y   (Z_1)^2]\}^{2} } \leq \frac{49}{c_3^\prime \ushort{f}^2 } \frac{J^2 }{n }   \   \text{ for $n$ sufficiently large} .
\end{equation*}Thus, the desired result arises as the right-hand side is clearly $ \lowmathcal{o} (1) $ by Assumption \ref{ass:sjn}.

To prove the first part of 3., $\E(\mathcal{U}^2) / \sigma^2 \rightarrow 1$, write $\nu_k : = \sum_{l=1}^k \binom{k}{l} (-1)^{k-l} \zeta_l \geq 0 $ for $k=1,\dots,s$, where $\zeta_l$ denotes the covariance between $\lowmathcal{h}_y  \left( Z_{(\boldsymbol{\iota})}  \right)$ and $\lowmathcal{h}_y  \left( Z_{(\boldsymbol{\iota}^\prime)}   \right)$ when the index vectors $(\boldsymbol{\iota} , \boldsymbol{\iota}^\prime)$ have $l$ elements in common. Note that $\nu_1 =  \zeta_1 = \E [ g_y   (Z_1)^2  ] =  n \sigma^2 /s^2$ and $ \zeta_s = \E [ \lowmathcal{h}_y  \left( Z_{1} ,   \dots,  Z_{s}  \right)^2   ]$. One the one hand, if  $\nu_k = 0$ for all $k \geq 2$, it follows from \citet[Theorem 6.1, Eq.\ (6.3)]{vit92} that $\E (\mathcal{U}^2 ) = \nu_1 s^2/n = \sigma^2$. One the other hand, if $\nu_k >0$ for some $k \geq 2$, Theorem 6.2 in \cite{vit92} and Cauchy-Schwartz inequality imply that $\sigma^2     \leq  \E (\mathcal{U}^2 ) \leq   \sigma^2    +n^{2(\beta-1)} \zeta_s \leq   \sigma^2    + n^{2(\beta-1)}  \|\mathcal{T}_x (y)\|_2^2 J^2$. So, the desired result emerges from part 1.\ of this lemma and Assumption \ref{ass:sjn}.(ii). Finally, the second part of  3.\ can be derived from \citet[Theorem 11.2]{vdVaart98}. \qed


\paragraph{Proof of Theorem \ref{thm:main}.} To prove Eq.\ (\ref{eq:asympf}), first, note that $ \| \mathcal{T}_x (y) \|_2 \sqrt{J}  \| \hat{\boldsymbol{\theta}}_x   -  \tilde{\boldsymbol{\theta}}_x   \|_2^2 / \sigma = \lowmathcal{o}_p(1)$ due to Assumption \ref{ass:sjn}, Lemma \ref{le:approx1}, and the condition $\beta > (1 + 2 \lowmathcal{c}_b)^{-1}$. Then, the desired asymptotic distribution can be derived by applying the following results to the term $\mathcal{U}$ from Lemma \ref{le:approx2}: parts 2.\ and 3.\ of Lemma \ref{lem:norm}, Slutsky's lemma, and Lyapunov's central limit theorem.
Finally, the consequence arises from Lemma \ref{le:approx1}, Eqs.\ (\ref{eq:approxinfo}) and (\ref{eq:sigmabound}), and the inequality \begin{equation*}
  \left|       \tilde{f} ( y  ;   \tilde{\boldsymbol{\theta}}_x )   -   \tilde f ( y ; \boldsymbol{\theta}_x   )    \right| \leq  \| \mathcal{T}_x (y) \|_2     \left\| \E  \left( \tilde{\boldsymbol \mu}_x \right) - \boldsymbol{\mu}_x   \right\|_2 +   \mathcal{O}  \left( \| \mathcal{T}_x (y) \|_2 \sqrt{J}   \| \tilde{\boldsymbol{\theta}}_x   - \boldsymbol{\theta}_x    \|_2^2   \right) 
\end{equation*}that can be obtained by arguments similar to those used in Eq.\ (\ref{eq:approxtay}). Note that combining together the requirement $ \gamma < \beta(1 +  2 \lowmathcal{c}_b) -1$ with Eq.\ (\ref{eq:sigmabound}) yields $ \| \mathcal{T}_x (y) \|_2 \sqrt{J}  / ( \sigma s^{\lowmathcal{c}_b} ) = \lowmathcal{o}(1)$. Moreover, we have that $ \| \mathcal{T}_x (y) \|_2 \sqrt{J}  \| \tilde{\boldsymbol{\theta}}_x   -  {\boldsymbol{\theta}}_x   \|_2^2 / \sigma = \lowmathcal{o}(1)$. \qed

\paragraph{Proof of Corollary \ref{cor:asympcov}.} It follows from Theorem \ref{thm:main} and Slutzky's lemma. \qed

\paragraph{Proof of Theorem \ref{thm:se}.} Considering the unfeasible estimator  \begin{equation*}
\tilde\sigma^2 =  \frac{n - D_{\sigma}}{D_{\sigma} } \binom{n}{D_{\sigma}}^{-1} \sum_{ \boldsymbol{\iota} \in   \mathscr{I}_{n,n-D_{\sigma}} } \left[    \mathcal{T}_x (y)    \left( \tilde{\boldsymbol \mu}_{x,\boldsymbol{\iota}}    -  \tilde{\boldsymbol \mu}_x   \right)  \right]^2  ,
\end{equation*}the desired result can be established if we show that \begin{equation}		\label{eq:provesig}
 \frac{1}{\sigma^2}  \left| \hat\sigma^2  - \tilde\sigma^2  \right|  = \lowmathcal{o}_p (1)   \  \text{and} \  \frac{\tilde\sigma^2 }{\sigma^2}  - 1  = \lowmathcal{o}_p (1)  .
\end{equation}To prove the first result, note that \begin{eqnarray*}
&  & \left| \hat\sigma^2  - \tilde\sigma^2  \right|  \\ 
&  & \  \  = \  \left| \frac{n - D_{\sigma}}{D_{\sigma} } \binom{n}{D_{\sigma}}^{-1} \sum_{ \boldsymbol{\iota} } \left\{  \left[  \hat{\mathcal{T}}_x (y)    - \mathcal{T}_x (y)   \right] \left( \tilde{\boldsymbol \mu}_{x,\boldsymbol{\iota}}    -  \tilde{\boldsymbol \mu}_x   \right)  \right\}  \left\{   \left[  \hat{\mathcal{T}}_x (y)    + \mathcal{T}_x (y)   \right]   \left( \tilde{\boldsymbol \mu}_{x,\boldsymbol{\iota}}    -  \tilde{\boldsymbol \mu}_x   \right)  \right\}  \right|  \\
 & & \quad \quad  \quad  \leq  \    \frac{n - D_{\sigma}}{D_{\sigma} }  \left\|  \hat{\mathcal{T}}_x (y)    - \mathcal{T}_x (y)    \right\|_2   \left[   \binom{n}{D_{\sigma}}^{-1} \sum_{ \boldsymbol{\iota} } \left\| \tilde{\boldsymbol \mu}_{x,\boldsymbol{\iota}}    -  \tilde{\boldsymbol \mu}_x   \right\|_2^2   \right] \left\|  \hat{\mathcal{T}}_x (y)    + \mathcal{T}_x (y)    \right\|_2  .
\end{eqnarray*}Observe that $\|  \hat{\mathcal{T}}_x (y)    -  \mathcal{T}_x (y)    \|_2 = \mathcal{O}_p ( \sqrt{J^3 s/n}    ) $ and that the inequality $\|  \hat{\mathcal{T}}_x (y)    + \mathcal{T}_x (y)    \|_2  \leq 3 \|  \mathcal{T}_x (y)    \|_2 $ holds w.p.a.1 due to Lemmas \ref{alemma:matrix}.(c) and \ref{le:approx1}.3. Note also that, by Jensen's inequality, we can bound \begin{multline*}
 \binom{n}{D_{\sigma}}^{-1} \sum_{ \boldsymbol{\iota} } \left\| \tilde{\boldsymbol \mu}_{x,\boldsymbol{\iota}}    -  \tilde{\boldsymbol \mu}_x   \right\|_2^2   \ \leq \    \binom{n}{D_{\sigma}}^{-1} \sum_{ \boldsymbol{\iota} } \left\| \tilde{\boldsymbol \mu}_{x,\boldsymbol{\iota}}    -  \E (\tilde{\boldsymbol \mu}_x )  \right\|_2^2   \ +  \   \left\|  \tilde{\boldsymbol \mu}_x  -  \E (\tilde{\boldsymbol \mu}_x )  \right\|_2^2  \\
 + \  2  \left\|  \tilde{\boldsymbol \mu}_x  -  \E (\tilde{\boldsymbol \mu}_x )  \right\|_2\sqrt{   \binom{n}{D_{\sigma}}^{-1} \sum_{ \boldsymbol{\iota} } \left\| \tilde{\boldsymbol \mu}_{x,\boldsymbol{\iota}}    -  \E (\tilde{\boldsymbol \mu}_x )  \right\|_2^2 }  .
\end{multline*}Thus, the first result in Eq.\ (\ref{eq:provesig}) follows immediately by combing together Eq.\ (\ref{eq:sigmabound}), Lemma \ref{le:approx1}.1, and the inequalities \begin{equation*}
\E \left[ \left\| \tilde{\boldsymbol \mu}_{x,\ushort{\boldsymbol{\iota}}}    -  \E (\tilde{\boldsymbol \mu}_x )  \right\|_2^2      \right]   \leq J \max_{j=1,\dots,J}  \E \left\{ [ \tilde{\mu}_{x,\ushort{\boldsymbol{\iota}},j}  - E (\tilde{\mu}_{x,j}) ]^2 \right\}  \leq J \frac{s}{n-D_\sigma}        ,
\end{equation*}which can be easily derived by the arguments used in the proof of Lemma \ref{le:approx1}.1.

To establish ratio consistency of $\tilde\sigma^2$, which is the second result in Eq.\ (\ref{eq:provesig}), write \begin{equation*}
\tilde\sigma^2  \ =\  \frac{n - D_{\sigma}}{D_{\sigma} } \binom{n}{D_{\sigma}}^{-1} \sum_{ \boldsymbol{\iota} \in \mathscr{I}_{n,n-D_{\sigma}}   } \left(  \mathcal{U}_{\boldsymbol{\iota}}  -  \mathcal{U}   \right)^2  \  =  \  \frac{n - D_{\sigma}}{D_{\sigma} } \binom{n}{D_{\sigma}}^{-1} \sum_{ \boldsymbol{\iota} \in \mathscr{I}_{n,n-D_{\sigma}}   } \left(  \mathcal{T}_{1,\boldsymbol{\iota}}    +  \mathcal{T}_{2,\boldsymbol{\iota} }  +  \mathcal{T}_{3 }   \right)^2 ,
\end{equation*}where $\mathcal{T}_{1,\boldsymbol{\iota} }    =     \mathcal{U}_{\boldsymbol{\iota}} -  \mathcal{U}_{\boldsymbol{\iota}}^\circ $, $\mathcal{T}_{2,\boldsymbol{\iota} }     =    \mathcal{U}_{\boldsymbol{\iota}}^\circ -  \mathcal{U}^\circ$, $\mathcal{T}_{3}     =      \mathcal{U}^\circ  - \mathcal{U} $, \begin{eqnarray*}
 \mathcal{U}_{\boldsymbol{\iota}} &  :  =   &   \mathcal{T}_x (y)    \left[ \tilde{\boldsymbol \mu}_{x,\boldsymbol{\iota}}    -  \E (\tilde{\boldsymbol \mu}_x )  \right]   \ =  \ \binom{n - D_\sigma}{s}^{-1} \sum_{\tilde{\boldsymbol{\iota}} \in \check{\mathscr{I}}_{\boldsymbol{\iota},s} }  \lowmathcal{h}_y  \left( Z_{\tilde{\iota}_1} ,   \dots,  Z_{\tilde{\iota}_s}  \right)  ,  \  \text{and}  \\
 \mathcal{U}_{\boldsymbol{\iota}}^\circ   &   = &  \frac{s}{n - D_\sigma} \sum_{m=1}^{n - D_\sigma}  {g}_y   (Z_{\iota_m}) .
\end{eqnarray*}Observe that Lemma \ref{lem:norm} and Lyapounov's central limit theorem imply that $\sigma / \mathcal{U}^\circ = \mathcal{O}_p (1)$ and therefore $\mathcal{T}_{3} / \sigma    = \lowmathcal{o}_p (1)$. So, to derive the desired result, it suffices to show that \begin{eqnarray}
& & \E \left[ \left( \frac{\mathcal{T}_{1 ,\ushort{\boldsymbol{\iota}} } }{\sigma} \right)^2 \right]  \ = \ \lowmathcal{o} (1)  \ \ \text{and}    \label{eq:jack1}   \\   
&  & \frac{n - D_{\sigma}}{D_{\sigma} } \binom{n}{D_{\sigma}}^{-1} \sum_{ \boldsymbol{\iota} \in \mathscr{I}_{n,n-D_{\sigma}} } \frac{\mathcal{T}_{2,\boldsymbol{\iota} }^2}{\sigma^2}  \ \convp \ 1     \label{eq:jack2}
\end{eqnarray}because the cross terms can be bounded from above by Cauchy-Schwarz inequality; for instance, \begin{equation*}
\left| \binom{n}{D_{\sigma}}^{-1} \sum_{ \boldsymbol{\iota} \in \mathscr{I}_{n,n-D_{\sigma}}   } \frac{\mathcal{T}_{1 ,{\boldsymbol{\iota}}} \mathcal{T}_{2 ,{\boldsymbol{\iota}} } } {\sigma^2}  \right|  \leq  \left[ \binom{n}{D_{\sigma}}^{-1} \sum_{ \boldsymbol{\iota} \in \mathscr{I}_{n,n-D_{\sigma}}   } \frac{\mathcal{T}_{1 ,{\boldsymbol{\iota}}}^2 } {\sigma^2}    \right]^{1/2} \left[ \binom{n}{D_{\sigma}}^{-1} \sum_{ \boldsymbol{\iota} \in \mathscr{I}_{n,n-D_{\sigma}}   } \frac{\mathcal{T}_{2 ,{\boldsymbol{\iota}}}^2 } {\sigma^2}    \right]^{1/2} .
\end{equation*}To prove the result in Eq.\ (\ref{eq:jack1}), denote $\sigma^{2\prime}  : =\E (   \mathcal{U}_{\ushort{\boldsymbol{\iota}}}^\circ  )   = n  \sigma^2 / ( n - D_\sigma  ) $ and write \begin{equation*}
\E \left[ \left( \frac{\mathcal{T}_{1 , \ushort{\boldsymbol{\iota}} } }{\sigma} \right)^2 \right]  = \frac{n - D_\sigma}{n} \E \left[ \left( \frac{  \mathcal{U}_{ \ushort{\boldsymbol{\iota}} } -  \mathcal{U}_{ \ushort{\boldsymbol{\iota}} }^\circ  }{\sigma^\prime} \right)^2 \right]  =  \frac{n - D_\sigma}{n} \left[ \frac{\E \left(   \mathcal{U}_{ \ushort{\boldsymbol{\iota}} }  \right)}{\sigma^{2\prime} }  -  1  \right]   . 
\end{equation*}Then, note that the utmost right-hand side is clearly $\lowmathcal{o} (1)$ by the same arguments as the ones used in the proof of Lemma \ref{lem:norm}.3.  To prove the result in Eq.\ (\ref{eq:jack2}), letting $\bar{g}_y   =  \sum_{i=1}^n g_y (Z_i) /n$ and $\tilde{G}_i = g_y (Z_i)^2 / \E[ g_y (Z_1)^2 ]$,  write \begin{multline*}
\frac{n - D_{\sigma}}{D_{\sigma} } \binom{n}{D_{\sigma}}^{-1} \sum_{ \boldsymbol{\iota} \in \mathscr{I}_{n,n-D_{\sigma}} } \frac{\mathcal{T}_{2,\boldsymbol{\iota} }^2}{\sigma^2} \  =  \   \frac{1}{\E[ g_y (Z_1)^2 ](n-1)} \sum_{i=1}^n \left[  g_y (Z_i)  - \bar{g}_y  \right]^2 \\
 =  \      \frac{n}{(n-1)} \left[ \left( \frac{1}{n} \sum_{i=1}^n   \tilde{G}_i   \right) +   \left( \frac{ \bar{g}_y}{\sqrt{\E[ g_y (Z_1)^2 ] }} \right)^2 \right] ;
\end{multline*}the first equality follows from Eq.\ (1.6) in \cite{sw89}, while the second is straightforward. To complete the proof, observe that $(1/n)\sum_{i=1}^n  \tilde{G}_i    -  1 = \lowmathcal{o}_p (1)$ follows by Chebyshev's inequality and Lemma \ref{lem:norm}.2, while $\bar{g}_y / \sqrt{\E[ g_y (Z_1)^2 ] }  = \lowmathcal{o}_p (1)$ by the Law of large numbers. \qed


\subsection{Proof of Auxiliary Lemma \ref{alemma:matrix}}	\label{sub:palemma}

\paragraph{(a)}   The first result follows immediately by orthonormality of the Legendre polynomials; note that $\int_0^1 [ v^\tau \boldsymbol{\phi}(t)]^2 dt = \| v \|_2 $. The second and third results can be obtained from Rayleigh's quotient and the fact that the matrix norm induced by the Euclidean norm is equal to the spectral norm. The fourth result follows by Cauchy–Schwarz inequality.

\paragraph{(b)}   To prove the first inequality, write \begin{equation*}
\mathcal H_{f} ( y ; {\boldsymbol{\theta}} )  = \tilde{f}( y ; \boldsymbol{\theta} )   \left\{ \left[ \boldsymbol{\phi} (y) -  \int\boldsymbol{\phi} (t) \tilde{f}( t ; \boldsymbol{\theta} ) dt \right] \left[ \boldsymbol{\phi} (y) -  \int\boldsymbol{\phi} (t) \tilde{f}( t ; \boldsymbol{\theta} ) dt \right]^\tau  -    \mathcal{V} (   \boldsymbol{\theta} )  \right\}
\end{equation*}and note that $\| \cdot \|_2$ is sub-multiplicative as a matrix norm. To derive the second, write \begin{eqnarray*}
& &   \mathcal{H}_{\psi}  ( \boldsymbol{\theta} )  \\
& & \   =  \   \tilde{f} ( y ; \tilde{\boldsymbol{\theta}}_x )^{-1}   \int \mathcal{T}_x (y) \boldsymbol{\phi} (t) \left[ \boldsymbol{\phi} (t) -  \int\boldsymbol{\phi} (\cdot ) \tilde{f}( \cdot ; \boldsymbol{\theta} )  \right] \left[ \boldsymbol{\phi} (t) -  \int\boldsymbol{\phi} (\cdot ) \tilde{f}(\cdot ; \boldsymbol{\theta} )  \right]^\tau   \tilde{f}( t ; \boldsymbol{\theta} )      d t   \\
& & \quad \quad  - \ \tilde{f} ( y ; \tilde{\boldsymbol{\theta}}_x )^{-1}   \int  \mathcal{T}_x (y) \boldsymbol{\phi} (t)\tilde{f}( t ; \boldsymbol{\theta} )      d t  \times  \mathcal{V} ( \boldsymbol{\theta} )  
\end{eqnarray*}and apply Cauchy–Schwarz inequality, combined with part (a).

\paragraph{(c)}   Pick any $v \in \mathbb{R}^J$ and note that the mean value theorem implies that there is $\check{\boldsymbol{\theta}}_v \in \Theta_n$ in the segment between $\boldsymbol{\theta}$ and $\boldsymbol{\theta}^\prime$ such that \begin{equation*}
\left[   \tilde{\mathcal{T}} (y ; \boldsymbol{\theta} )   -   \tilde{\mathcal{T}} (y ; \boldsymbol{\theta}^\prime) \right] v = \sum_{j=1}^J ( \theta_j    - \theta_{j}^\prime   ) \frac{\partial \tilde{\mathcal{T}} (y ; \check{\boldsymbol{\theta}}_v ) }{\partial \theta_{j} }  v ,
\end{equation*}where \begin{eqnarray*}
\frac{\partial \tilde{\mathcal{T}} (y ; \boldsymbol{\theta}) }{\partial \theta_{j^\prime} }  &  = & \tilde{f} \left( y ; \boldsymbol{\theta} \right) \Big\{ \left[ {\phi}_j (y) -  \int{\phi}_j ( \cdot ) \tilde{f}( \cdot ; \boldsymbol{\theta} )  \right]    \left[ \boldsymbol{\phi} (y) -  \int\boldsymbol{\phi} (  \cdot ) \tilde{f}( \cdot ; \boldsymbol{\theta} )  \right]^\tau  \\
&   &   \    \  -  \  \int \left[ \phi_j (t) -  \int {\phi_j} (\cdot ) \tilde{f}( \cdot ; \boldsymbol{\theta} )  \right] \left[ \boldsymbol{\phi} (t) -  \int\boldsymbol{\phi} ( \cdot ) \tilde{f}( \cdot ; \boldsymbol{\theta} )  \right]^\tau   \tilde{f}(  t ; \boldsymbol{\theta} ) d t \Big\}  \mathcal{V} (  \boldsymbol{\theta}  )^{-1} \\
&   &   \    \  - \  \tilde{f}( y ; \boldsymbol{\theta} )   \left[ \boldsymbol{\phi} (y) -  \int\boldsymbol{\phi} ( \cdot ) \tilde{f}( \cdot ; \boldsymbol{\theta} )   \right]^\tau  \mathcal{V} (  \boldsymbol{\theta}  )^{-1} \left(  \frac{ \partial  \mathcal{V} (  \boldsymbol{\theta}  ) }{  \partial \theta_j  }  \right)   \mathcal{V} (  \boldsymbol{\theta}  )^{-1} 
\end{eqnarray*}and \begin{eqnarray*}
 \frac{ \partial  \mathcal{V} (  \boldsymbol{\theta}  ) }{  \partial \theta_j  }   &  =  & \int \boldsymbol{\phi} (t) \boldsymbol{\phi} (t)^\tau   \left[ {\phi}_j (t) -  \int {\phi}_j (t^\prime) \tilde{f}( t^\prime ; \boldsymbol{\theta} ) d t^\prime \right] \tilde{f}( t ; \boldsymbol{\theta} )  dt  \\
 &  &   - 2 \left\{ \int \boldsymbol{\phi} (t)   \left[ {\phi}_j (t) -  \int {\phi}_j (t^\prime) \tilde{f}( t^\prime ; \boldsymbol{\theta} ) d t^\prime \right] \tilde{f}( t ; \boldsymbol{\theta} ) dt  \right\}  \left\{ \int \boldsymbol{\phi} (t)  \tilde{f}( t ; \boldsymbol{\theta} ) dt  \right\}^\tau .
\end{eqnarray*}Observe that \begin{multline*}
\sum_{j=1}^J ( \theta_j    - \theta_{j}^\prime   )    \frac{ \partial  \mathcal{V} (  \check{\boldsymbol{\theta}}_v  ) }{  \partial \theta_j  } \  = \ \int \boldsymbol{\phi} (t) \boldsymbol{\phi} (t)^\tau   \left\{ \left[\boldsymbol{\phi} (t) -  \int \boldsymbol{\phi} (\cdot) \tilde{f}( \cdot; \check{\boldsymbol{\theta}}_v   )  \right]^\tau \left(  \boldsymbol{\theta} - \boldsymbol{\theta}^\prime \right) \right\} \tilde{f}( t ; \check{\boldsymbol{\theta}}_v   )  dt \\
- \ 2 \mathcal{V} ( \check{\boldsymbol{\theta}}_v  ) \left(  \boldsymbol{\theta} - \boldsymbol{\theta}^\prime \right)   \left\{ \int \boldsymbol{\phi} (t)  \tilde{f}( t ;  \check{\boldsymbol{\theta}}_v    ) dt  \right\}^\tau .
\end{multline*}Considering now the terms on the right-hand side, for any $w \in \mathbb{R}^J$, note that we can bound \begin{equation*}
\left|   \int [ w^\tau \boldsymbol{\phi} (t)  ]^2  \left\{ \left[\boldsymbol{\phi} (t) -  \int \boldsymbol{\phi} (\cdot) \tilde{f}( \cdot; \check{\boldsymbol{\theta}}_v   )  \right]^\tau \left(  \boldsymbol{\theta} - \boldsymbol{\theta}^\prime \right) \right\} \tilde{f}( t ; \check{\boldsymbol{\theta}}_v   )  dt  \right| 
\leq   2 \bar{f}  \sqrt{J} \left\| \boldsymbol{\theta} - \boldsymbol{\theta}^\prime  \right\|_2 \| w\|_2^2 
\end{equation*}and $\| \mathcal{V} ( \check{\boldsymbol{\theta}}_v  ) (  \boldsymbol{\theta} - \boldsymbol{\theta}^\prime )   \{ \int \boldsymbol{\phi} (t)  \tilde{f}( t ;  \check{\boldsymbol{\theta}}_v    ) dt  \}^\tau    \|_2   \leq  2 \bar{f} \sqrt{J} \|  \boldsymbol{\theta} - \boldsymbol{\theta}^\prime \|_2$. The desired result can then be obtained from the inequality\begin{multline*}
 \left|    \left[   \tilde{\mathcal{T}} (y ; \boldsymbol{\theta} )   -   \tilde{\mathcal{T}} (y ; \boldsymbol{\theta}^\prime ) \right] v  \right| \   \leq  \  4 \bar{f} J \| \boldsymbol{\theta} -  \boldsymbol{\theta}^\prime \|_2  \| v\|_2   +  4 J \| \boldsymbol{\theta} -  \boldsymbol{\theta}^\prime \|_2  \| \mathcal{V} (   \check{\boldsymbol{\theta}}_v     )^{-1}  v \|_2    \\
 +  \  \left\| \tilde{\mathcal{T}} (y ;   \check{\boldsymbol{\theta}}_v )   \right\|_2 \left( 2 \bar{f}  \sqrt{J} \left\| \boldsymbol{\theta} - \boldsymbol{\theta}^\prime  \right\|_2   + 4 \bar{f} \sqrt{J} \|  \boldsymbol{\theta} - \boldsymbol{\theta}^\prime \|_2 \right)    \| \mathcal{V} (  \check{\boldsymbol{\theta}}_v    )^{-1}  v \|_2    
\end{multline*}together with part (a) of this auxiliary lemma, and the fact that $\| w  \|_2 \leq \sup_{\| v\|_2 = 1} | w^\tau v  |$ for all $w \in \mathbb{R}^J$.		\qed	



\section{Monte Carlo experiments: Additional results}		\label{sec:addmc}

\begin{figure}[ttt!]

\caption{Performance of $\hat{f}(\cdot|x)$ when $n=1,000$ from Monte Carlo experiments.}

\label{fig:mcsupple}

\begin{center}

\includegraphics[scale=.425]{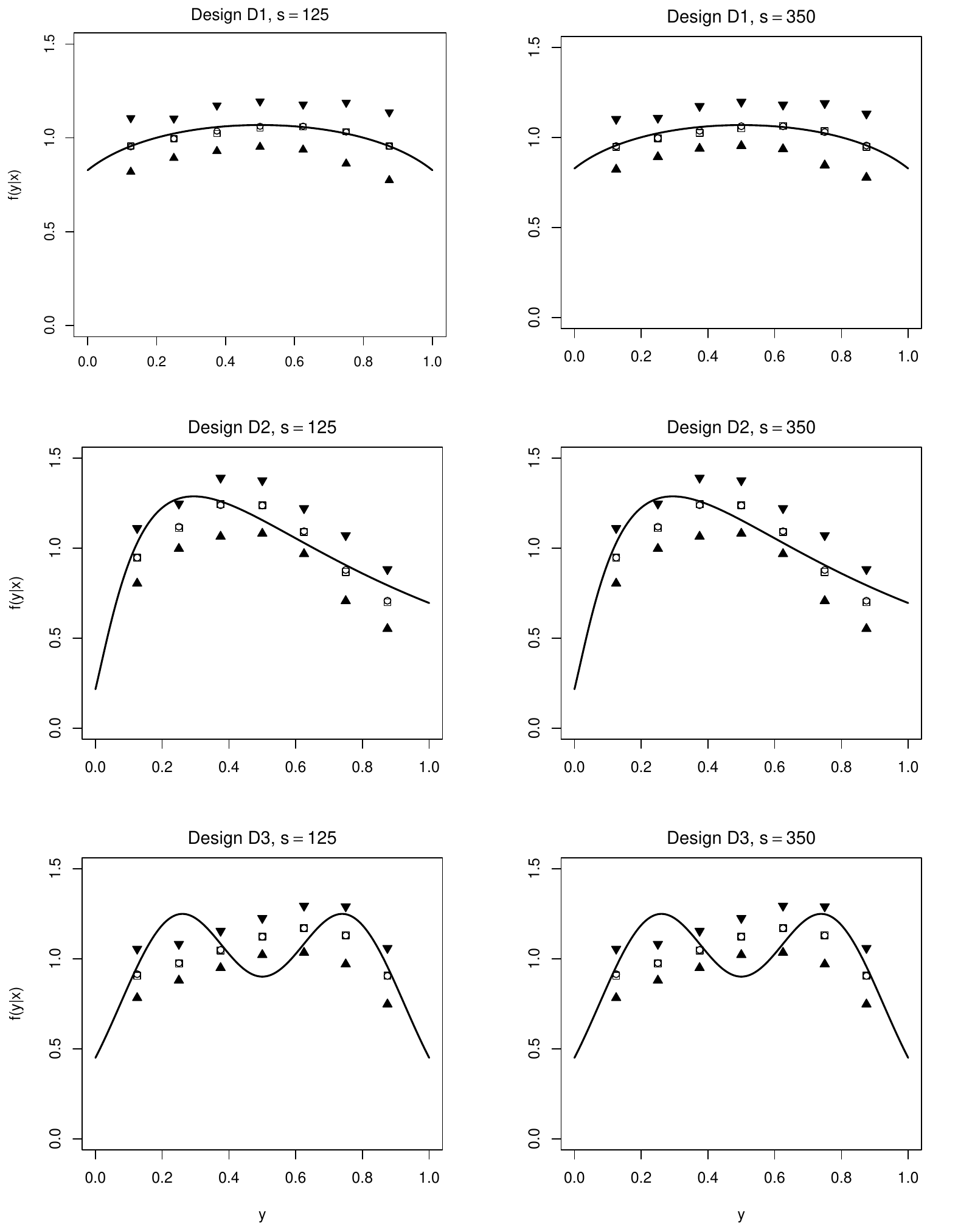}

\end{center}
{\raggedright\footnotesize{Solid line is the true conditional density with $x = (1/2) \mathbf{1}_4$. Circles and squares are the obtained mean and median of $\hat{f}( y| x)$ for the corresponding values of $y$. Black up- and down-pointing triangles are the obtained $10^\text{th}$ and $90^\text{th}$ percentiles of $\hat{f}( y| x)$, respectively.}}

\end{figure}

This section provides additional results from Monte Carlo simulations. The results are presented in Table \ref{table:mcsupple} and Figure \ref{fig:mcsupple} in the next pages, which can be compared with Table \ref{table:perf} and Figure \ref{fig:mc}, respectively, of the main text.

The simulations were conducted in essentially the same manner as in Section \ref{sec:mc}, differing only in the choice of $s$ to perform a robustness check. Specifically, here I used $s \in \{ 75 ,  200 \}$ and $s \in \{ 125, 375 \}$ as subsample sizes for $n=500$ and $n = 1000$, respectively. The performance of the proposed estimator $\hat{f} (\cdot | x) $ remains essentially unaffected. Results from the suggested standard error formula have been omitted, as its performance is suboptimal for low values of $s$, based on a few preliminary simulations that are not reported here.

\clearpage

\begin{table}[ttt!]
\renewcommand{\arraystretch}{1.25}
	\caption{Bias, std.\ deviation, and MISE of $\hat{f}(\cdot| x)$ from simulations.} \label{table:mcsupple}
	\begin{center}
		\begin{scriptsize}
			\begin{tabular}{!{\vrule width 1.5pt}       l  |  l  |  l   ||   r  r |  r r    ||  r r  | r  r  !{\vrule width 1.5pt}}
				
				\thickhline
				
				\multicolumn{1}{!{\vrule width 1.5pt} l |}{\multirow{3}{*}{Design}}  &  \multicolumn{1}{l |}{\multirow{3}{*}{$y$}} & \multicolumn{1}{l ||}{\multirow{3}{*}{$f (y | x)$ }}  &  \multicolumn{4}{c||}{$n=500$} &  \multicolumn{4}{c !{\vrule width 1.5pt}}{$n=1,000$}  \\     \cline{4-11}

	\multicolumn{1}{!{\vrule width 1.5pt} c |}{} 		 & \multicolumn{1}{ c |}{}   &  &  \multicolumn{2}{c|}{GRF}    & \multicolumn{1}{c|}{\multirow{2}{*}{Kernel}} & \multicolumn{1}{c||}{\multirow{2}{*}{MLE}} & \multicolumn{2}{c|}{GRF} & \multicolumn{1}{c|}{\multirow{2}{*}{Kernel}}   & \multicolumn{1}{c!{\vrule width 1.5pt}}{\multirow{2}{*}{MLE}}   \\   \cline{4-5}  \cline{8-9}
				
				\multicolumn{1}{!{\vrule width 1.5pt} c |}{} 		 & \multicolumn{1}{ c |}{}   &  & \multicolumn{1}{ l  |}{s=75}  & \multicolumn{1}{l|}{s= 200}   & \multicolumn{1}{c|}{ } & \multicolumn{1}{c||}{ } &      \multicolumn{1}{ l  |}{s=125}  & \multicolumn{1}{l|}{s= 375}    &\multicolumn{1}{ c  |}{}    & \multicolumn{1}{c !{\vrule width 1.5pt}}{} 	 \\

\thickhline

D1	&	0.125	&	0.956	&	-0.011	&	-0.012	&	0.060	&	-0.009	&	-0.003	&	-0.004	&	0.080	&	-0.004	\\
	&	 	&	 	&	(0.16)	&	(0.16)	&	(1.13)	&	(0.06)	&	(0.11)	&	(0.12)	&	(1.03)	&	(0.04)	\\
	&	0.250	&	1.023	&	-0.033	&	-0.034	&	0.104	&	-0.002	&	-0.028	&	-0.027	&	0.041	&	-0.001	\\
	&	 	&	 	&	(0.11)	&	(0.11)	&	(1.23)	&	(0.04)	&	(0.08)	&	(0.08)	&	(0.99)	&	(0.03)	\\
	&	0.375	&	1.058	&	-0.019	&	-0.021	&	0.174	&	0.004	&	-0.022	&	-0.021	&	0.056	&	0.002	\\
	&	 	&	 	&	(0.13)	&	(0.13)	&	(1.31)	&	(0.04)	&	(0.09)	&	(0.09)	&	(1.01)	&	(0.03)	\\
	&	0.500	&	1.069	&	0.002	&	0.001	&	0.036	&	0.007	&	-0.007	&	-0.006	&	0.053	&	0.003	\\
	&	 	&	 	&	(0.13)	&	(0.13)	&	(1.20)	&	(0.04)	&	(0.09)	&	(0.09)	&	(1.03)	&	(0.03)	\\
	&	0.625	&	1.058	&	0.015	&	0.016	&	0.062	&	0.007	&	0.005	&	0.004	&	0.061	&	0.003	\\
	&	 	&	 	&	(0.13)	&	(0.13)	&	(1.21)	&	(0.04)	&	(0.09)	&	(0.10)	&	(1.05)	&	(0.03)	\\
	&	0.750	&	1.023	&	0.015	&	0.017	&	0.088	&	0.004	&	0.008	&	0.007	&	-0.032	&	0.002	\\
	&	 	&	 	&	(0.17)	&	(0.17)	&	(1.16)	&	(0.04)	&	(0.13)	&	(0.13)	&	(0.87)	&	(0.03)	\\
	&	0.875	&	0.956	&	-0.012	&	-0.010	&	0.005	&	-0.001	&	0.000	&	-0.001	&	0.057	&	0.000	\\
	&	 	&	 	&	(0.18)	&	(0.18)	&	(1.04)	&	(0.06)	&	(0.13)	&	(0.14)	&	(0.99)	&	(0.04)	\\  \cline{2-11}
	&	 \multicolumn{2}{l||}{\emph{MISE}} 	&	\emph{0.014}	&	\emph{0.014}	&	\emph{1.036}	&	\emph{0.001}	&	\emph{0.007}	&	\emph{0.008}	&	\emph{0.687}	&	\emph{0.001}	\\
	
\thickhline																					
D2	&	0.125	&	1.028	&	-0.086	&	-0.087	&	0.071	&	-0.017	&	-0.081	&	-0.081	&	-0.113	&	-0.011	\\
	&	 	&	 	&	(0.17)	&	(0.17)	&	(1.50)	&	(0.10)	&	(0.11)	&	(0.12)	&	(0.98)	&	(0.07)	\\
	&	0.250	&	1.275	&	-0.161	&	-0.161	&	0.117	&	-0.002	&	-0.157	&	-0.156	&	0.129	&	0.000	\\
	&	 	&	 	&	(0.14)	&	(0.14)	&	(1.37)	&	(0.07)	&	(0.09)	&	(0.10)	&	(1.14)	&	(0.05)	\\
	&	0.375	&	1.259	&	-0.022	&	-0.021	&	0.119	&	0.005	&	-0.024	&	-0.021	&	0.109	&	0.005	\\
	&	 	&	 	&	(0.16)	&	(0.16)	&	(1.44)	&	(0.05)	&	(0.12)	&	(0.12)	&	(1.17)	&	(0.04)	\\
	&	0.500	&	1.155	&	0.081	&	0.082	&	0.069	&	0.006	&	0.079	&	0.081	&	-0.030	&	0.005	\\
	&	 	&	 	&	(0.16)	&	(0.16)	&	(1.34)	&	(0.04)	&	(0.11)	&	(0.11)	&	(1.00)	&	(0.03)	\\
	&	0.625	&	1.029	&	0.062	&	0.063	&	0.184	&	0.006	&	0.064	&	0.065	&	0.003	&	0.004	\\
	&	 	&	 	&	(0.14)	&	(0.14)	&	(1.52)	&	(0.04)	&	(0.09)	&	(0.10)	&	(1.01)	&	(0.03)	\\
	&	0.750	&	0.906	&	-0.029	&	-0.030	&	0.006	&	0.004	&	-0.025	&	-0.027	&	0.084	&	0.002	\\
	&	 	&	 	&	(0.18)	&	(0.18)	&	(1.12)	&	(0.05)	&	(0.14)	&	(0.14)	&	(1.01)	&	(0.04)	\\
	&	0.875	&	0.794	&	-0.086	&	-0.087	&	0.145	&	0.003	&	-0.082	&	-0.086	&	0.132	&	0.001	\\
	&	 	&	 	&	(0.18)	&	(0.18)	&	(1.29)	&	(0.06)	&	(0.13)	&	(0.13)	&	(0.95)	&	(0.05)	\\ \cline{2-11}
	&	\multicolumn{2}{l||}{\emph{MISE}}	&	\emph{0.023}	&	\emph{0.023}	&	\emph{1.333}	&	\emph{0.002}	&	\emph{0.015}	&	\emph{0.015}	&	\emph{0.797}	&	\emph{0.001}	\\
\thickhline																					
D3	&	0.125	&	0.956	&	-0.049	&	-0.054	&	0.029	&	0.000	&	-0.043	&	-0.041	&	-0.007	&	0.000	\\
	&	 	&	 	&	(0.16)	&	(0.16)	&	(1.28)	&	(0.05)	&	(0.11)	&	(0.11)	&	(0.92)	&	(0.03)	\\
	&	0.250	&	1.247	&	-0.285	&	-0.286	&	0.009	&	0.005	&	-0.272	&	-0.271	&	0.034	&	0.001	\\
	&	 	&	 	&	(0.12)	&	(0.12)	&	(1.35)	&	(0.06)	&	(0.08)	&	(0.08)	&	(1.10)	&	(0.04)	\\
	&	0.375	&	1.083	&	-0.044	&	-0.044	&	0.031	&	-0.001	&	-0.034	&	-0.034	&	0.019	&	-0.001	\\
	&	 	&	 	&	(0.13)	&	(0.13)	&	(1.19)	&	(0.04)	&	(0.08)	&	(0.08)	&	(1.07)	&	(0.03)	\\
	&	0.500	&	0.901	&	0.218	&	0.220	&	0.043	&	-0.007	&	0.222	&	0.221	&	0.057	&	-0.003	\\
	&	 	&	 	&	(0.14)	&	(0.14)	&	(1.17)	&	(0.09)	&	(0.08)	&	(0.09)	&	(0.91)	&	(0.06)	\\
	&	0.625	&	1.083	&	0.093	&	0.097	&	0.228	&	-0.001	&	0.087	&	0.087	&	0.103	&	-0.001	\\
	&	 	&	 	&	(0.14)	&	(0.14)	&	(1.34)	&	(0.04)	&	(0.10)	&	(0.10)	&	(1.06)	&	(0.03)	\\
	&	0.750	&	1.247	&	-0.098	&	-0.096	&	0.045	&	0.005	&	-0.118	&	-0.118	&	0.000	&	0.001	\\
	&	 	&	 	&	(0.20)	&	(0.20)	&	(1.35)	&	(0.06)	&	(0.12)	&	(0.13)	&	(1.08)	&	(0.04)	\\
	&	0.875	&	0.956	&	-0.045	&	-0.045	&	0.046	&	0.000	&	-0.050	&	-0.051	&	0.094	&	0.000	\\
	&	 	&	 	&	(0.19)	&	(0.19)	&	(1.13)	&	(0.05)	&	(0.12)	&	(0.12)	&	(1.01)	&	(0.03)	\\ \cline{2-11}
	&	\multicolumn{2}{l||}{\emph{MISE}}		&	\emph{0.035}	&	\emph{0.036}	&	\emph{1.078}	&	\emph{0.002}	&	\emph{0.025}	&	\emph{0.025}	&	\emph{0.775}	&	\emph{0.001}	\\
\thickhline

			\end{tabular}
		\end{scriptsize}
	\end{center}

{\raggedright \footnotesize{Bias and std.\ deviation are reported in normal fonts under the columns `GRF', `Kernel,' and `MLE;' std.\ deviation is in parentheses. MISE over [0.15,0.85] is reported in \emph{italics}.}}

\end{table}

\FloatBarrier

\section{Empirical illustration}		\label{sec:ei}

In this section, I apply the proposed estimator $\hat{f}(\cdot |x)$ to a real-world dataset of timber auctions from the U.S.\ Forest Service. Specifically, I consider the dataset used in \cite{liuluo17} and refer to this article for a detailed description, as well as to the references cited therein.

The considered dataset contains a sample of first-price sealed-bid auctions, each having a timber track as auctioned object. In each auction, we observe the bids from all the participants (in nominal dollars) and several auction-specific covariates, from which I consider two of them in this application: $vol$: estimated volume of timber in thousand board-feet (MBF); $AppVal$:  appraisal value per MBF. I consider only auctions that have two participants, giving a total of 197 auctions and 394 bids. On this subsample, the sample averages of $vol$ and $AppVal$ are $1705.73$ and $35.98$, respectively.

Next, I provide nonparametric estimates of the conditional density of bids given $( vol , AppVal) = (  \exp(6.4)  , \exp(3.2) )$ and explain how these estimates have been obtained. Specifically, the procedure can be described in four steps. \begin{enumerate}

\item  Letting $B_{p,m}$ denote the bid (in thousands of dollars) of participant $p$ in the $m$th auction and letting $(vol_m  , AppVal_m ) $ be the corresponding auction-specific covariates, apply the logarithmic transformation to each observation and write \begin{equation*}
\left( lB_{1,m} , lB_{2,m}  , lvol_m  , lAppVal_m  \right) = \left( \log(B_{1,m}) , \log(B_{2,m})  , \log(vol_m)  , \log(AppVal_m)  \right) .
\end{equation*}

\item Letting $[ \ushort{b} (x) , \bar{b}(x) ]$ denote the support of $lB$ conditional on $( lvol , lAppVal) = x$, estimate $\ushort{b} ( x )$ by simply setting $\hat{\ushort{b}} = \min\{  lB_{p,m}   :  p =1 ,2, \ m = 1,\dots, 197  \}$ for all $x$. Estimate $\bar{b}( \cdot ) $ using \cite{kt93}'s piecewise-polynomial estimator and denote it by $\hat{\bar{b}}( \cdot )$. Compute and store the standarized bids as follows: \begin{equation*}
 \widetilde{lB}_{p,m}  = \frac{ {lB}_{p,m}  - \hat{\ushort{b}}  }{ \hat{\bar{b}}(  vol_m  , AppVal_m    )  - \hat{\ushort{b}}  }      \in [0 , 1] .
\end{equation*}

\item To simplify the double indices, with a slight abuse of notation, redefine \begin{eqnarray*}
  \widetilde{lB}_{i}   & \leftarrow  &  \left\{ \begin{array}{ll}
 \widetilde{lB}_{1,i}   &  \text{if} \ i \leq 197 , \\
  \widetilde{lB}_{2,i-197}   &  \text{if} \ i > 197
\end{array}  \right.  \ \text{and}   \\ 
 (  lvol_i  , lAppVal_i    ) & \leftarrow & \left\{ \begin{array}{ll}
 (  lvol_i  , lAppVal_i    )   &  \text{if} \ i \leq 197 , \\
  (  lvol_{i-197}  , lAppVal_{i-197}    )  &  \text{if} \ i > 197
  \end{array}   \right.
\end{eqnarray*}for $i = 1,\dots, 397$. Then, setting $x = ( 6.4 ,  3.2)$, apply Algorithm 1 to the sample \begin{equation*}
\{ \left(  \widetilde{lB}_{i} ,   (  lvol_i  , lAppVal_i ) \right)  : i = 1,\dots,397  \}
\end{equation*}and store the estimated function as $\hat{f} (\cdot | x)$.

\item Using the transformation formula, compute the desired estimate as \begin{equation*}
\hat{f}_{B| vol , AppVal } \left( y |    \exp(6.4)  , \exp(3.2)   \right) = \frac{1}{y\left( \hat{\bar{b}}( 6.4 ,  3.2)  -  \hat{\ushort{b}} \right) } \hat{f}  \left( \frac{\log(y) -  \hat{\ushort{b}}  }{\hat{\bar{b}}( 6.4 ,  3.2)  -  \hat{\ushort{b}}}   \middle| x  \right).
\end{equation*}for $y \in [ \exp(\hat{\ushort{b}}) , \exp[ \hat{\bar{b}}( 6.4 ,  3.2) ]   ]$.

\end{enumerate}

I remark that, in the first step, we apply a logarithmic transformation to deal with extreme values and to help normalize the distribution. This becomes particularly helpful to compute \cite{kt93}'s upper boundary estimator in the second step. To implement such an estimator, I have divided the support of $( lvol , lAppVal) $ into four rectangles, having \begin{equation*}
\left( \frac{\max\{lvol_l \} - \min\{lvol_l \}}{2} ,  \frac{\max\{lAppVal_l \} - \min\{lAppVal_l \}}{2} \right) 
\end{equation*}as common vertex, and I have used a quadratic polynomial over each of them. For estimating $\ushort{b}(\cdot)$, I implicitly assume that the lower boundary of the latent distribution of private values does not depend on the covariates, so neither does $\ushort{b}(\cdot)$; hence, I simply suggest using $\hat{\ushort{b}} = \min\{  lB_{p,m} \}$ as an estimator. In the third step, I have employed Algorithm 1 with the following tuning parameters: $J=6$, $N=2240$, $\ushort{k} = 10$, $\ushort{\alpha} = 0.05$, and initial parent node $P = [4.5,8.5] \times [1.5 , 4.5]$. Two subsample sizes have been considered: $s = 75$ and $s=180$. Regarding the fourth step, the estimated boundaries for this application are $  \exp(\hat{\ushort{b}}) \approx 2.1$ and $  \exp[ \hat{\bar{b}}( 6.4 ,  3.2) ]\approx 36.8$.

The estimated conditional densities are depicted in Figure \ref{fig:empapp} below. The figure on the left shows the estimated conditional densities of standardized bids (third step), while the one on the right presents the conditional densities of bids (in thousands of dollars) obtained in the fourth step. The solid lines correspond to $s = 75$, while the dashed ones to $s=180$. The obtained results indicate that the conditional density of bids exhibits asymmetry and a single peak, with the majority of its mass concentrated on the left side of the distribution. The solid line on the left figure suggests that the conditional density of bids might be log-normal. 

\begin{figure}[ttt]

\caption{Estimated conditional densities given $( vol , AppVal) = (  \exp(6.4)  , \exp(3.2) )$.}

\label{fig:empapp}

\vspace{.1in}

\begin{center}

\includegraphics[scale=.325]{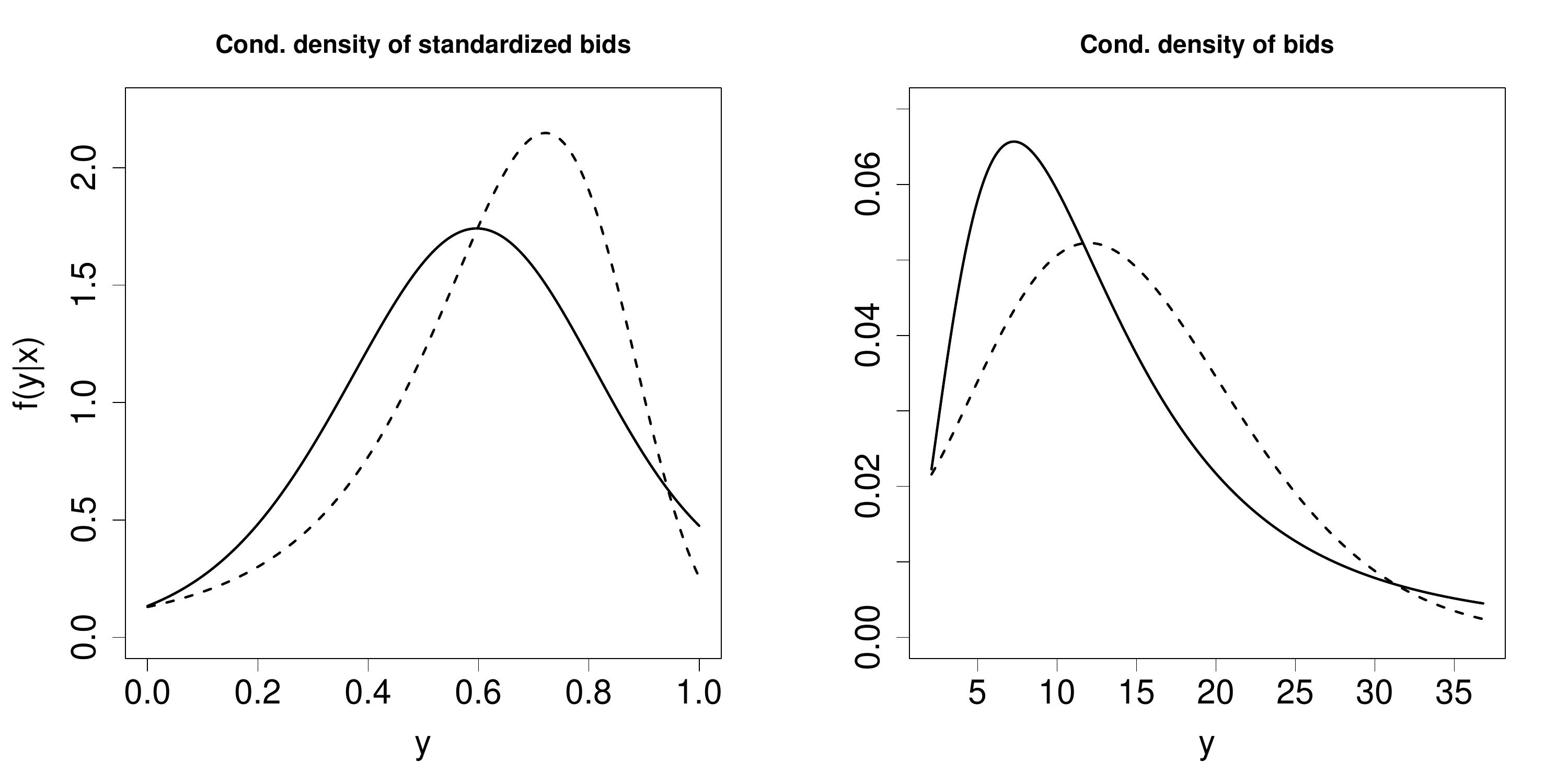}

\end{center}

{\raggedright\footnotesize{Solid line corresponds to $s = 75$, dashed line to $s = 180$.}}

\end{figure}

\clearpage

\bibliographystyle{Chicago}

\bibliography{bib_zin_cden_grf}

\end{document}